\documentclass[12pt,twoside]{report} 
\usepackage{latexsym}
  
\begin{document}
    \setcounter{chapter}{7}


\chapter[Relativistic QM and QFT]{Relativistic Quantum Mechanics and Quantum Field Theory\footnote{Chapter 8. 
of the book {\it Applied Bohmian Mechanics: From Nanoscale Systems to Cosmology}, 2nd edition,
edited by X. Oriols and J. Mompart (Jenny Stanford Publishing, 2019).
}
\label{ch_nikolic}}


\vspace{1.0cm}

\Large
{\bf Hrvoje Nikoli\'c}
\normalsize

\vspace{1.0cm}

\noindent
{\it Theoretical Physics Division, Rudjer Bo\v{s}kovi\'c Institute \\
P.O.B. 180, HR-10002 Zagreb, Croatia}

\vspace{0.5 cm}

\noindent
e-mail: hnikolic@irb.hr

\vspace{0.5 cm}

\begin{center}
{\bf Abstract} 
\end{center}
A general formulation of classical relativistic particle mechanics is presented,
with an emphasis on the fact that superluminal velocities and nonlocal interactions
are compatible with relativity. Then a manifestly relativistic-covariant formulation 
of relativistic quantum mechanics (QM) of fixed number of particles (with or without spin) is presented,
based on many-time wave functions and the spacetime probabilistic interpretation.
These results are used to formulate the Bohmian interpretation of relativistic QM in a
manifestly relativistic-covariant form. The results are also generalized to quantum field theory (QFT),
where quantum states are represented by
wave functions depending on an infinite number of spacetime coordinates.
The corresponding Bohmian interpretation of QFT describes an infinite number
of particle trajectories. Even though the particle trajectories are continuous, 
the appearance of creation and destruction of a finite number of
particles results from quantum theory of measurements
describing entanglement with particle detectors.

\setcounter{secnumdepth}{3}  
\setcounter{tocdepth}{3}         

\tableofcontents


\newpage

\section{Introduction}\label{nikolic:sec_intro}

The following chapter somewhat differs from the previous ones, in the sense that
this chapter does not deal with an application to a specific practical physical problem.
Instead, the main goal of this chapter is to develop a {\em generalized formulation}
of Bohmian mechanics, such that effects of relativistic quantum mechanics 
(QM) and quantum field theory (QFT) can also be incorporated into it. 

Since this chapter deals with a general formulation of the theory, the practical
utility of it may not be obvious. Nevertheless, we believe that the results of this
chapter may lead to practical applications as well. 
For example, many physicists argue that the most practical result that emerged 
from the original Bohm reformulation of QM was the famous Bell theorem 
that revealed fundamental nonlocal nature of QM.\footnote{Some 
even argue that the Bell theorem is the most important discovery
in physics of the 20th century.} The Bell's result is valid independently
on validity of the Bohm reformulation, but to obtain this result Bell was significantly 
guided and inspired by the Bohm reformulation in which nonlocality of QM is particularly 
manifest.\footnote{Today many physicists still misinterpret the Bell theorem
as a proof that the Bohmian interpretation cannot be right. It cannot 
be overemphasized that just the opposite is true. Bell theorem proves that no
{\em local} hidden variable theory can be compatible with QM, so that any
hidden variable theory compatible with QM must necessarily be {\em nonlocal}.
Since the Bohmian interpretation is a nonlocal hidden variable theory, 
the Bell theorem gives further credit to it. Indeed, Bell himself
had a very positive opinion on the Bohmian interpretation and significantly
contributed to the popularization of it.}
In a similar way, although our primary motivation lying behind the results of this
chapter is to make the Bohmian formulation compatible with relativity and QFT,
this motivation led us to some new fundamental results on relativistic QM and QFT
valid even without the Bohmian interpretation. In our quest towards
relativistic Bohmian mechanics, as a byproduct we realize that even
non-Bohmian relativistic QM and QFT should be first made ``more relativistic''
than they are in the usual formulation, i.e., that time and space should be treated
more symmetrically. First, the usual single-time wave function should be generalized
to the many-time wave function, such that each particle has its own spacetime coordinate.
Second, $|\psi|^2$ should be reinterpreted as a probability density in spacetime, rather than that in space. 
Eventually, this byproduct may turn out to be even more useful than the
relativistic Bohmian formulation itself.\footnote{For example, the many-time
formalism with spacetime probability density can be used to avoid the black-hole
information paradox \cite{nikbh1,nikbh2}.} 

The primary motivation lying behind this chapter has very much to do with
nonlocality of QM. One of the most frequent questions related to nonlocality
is how can it be compatible with relativity? If entangled particles communicate
instantaneously, is it in contradiction with the relativistic rule that no information 
can travel faster than light?  Since communication instantaneous in
one Lorentz frame cannot be instantaneous in any other Lorentz frame, 
does it mean that there exists a preferred Lorentz frame which violates 
(the spirit of) relativity?
In this chapter we offer an answer to these and many other questions
regarding relativity, nonlocality, quantum mechanics, and Bohmian mechanics.
In particular, by developing the Bohmian interpretation of QFT,
we also explain how continuous particle trajectories can be made
compatible with phenomena of particle creation and destruction.

Of course, due to the lack of space, this chapter is not intended to be 
a general pedagogic introduction to relativistic QM and QFT. Instead, we assume
that the reader is already familiar with some basics of those, as well as 
with some basics of special relativity in classical mechanics. (A knowledge of some
basics of general relativity may also be useful, but is not necessary.)
With these assumptions, we can pay more attention to aspects that are
not widely known to experts in relativistic QM and QFT.

Our presentation is to a large extent based on the papers \cite{nikijqi,nikqft,niktorino,nikscalpot}.
Some results from the papers \cite{nik1,nik2,nik3,nik4,niktimeprob} are also used.

\section{Classical relativistic mechanics}\label{nikolic:sec_rel}

\subsection{Kinematics}

Our point of departure is a 4-dimensional spacetime with coordinates $x^{\mu}$,
$\mu=0,1,2,3$, and the Minkowski metric
$\eta_{\mu\nu}$, where $\eta_{00}=1$, $\eta_{ij}=-\delta_{ij}$, $\eta_{0i}=0$,
for $i=1,2,3$. We work in units in which the velocity of light is $c\equiv 1$.
At some places we also use the notation $x^{\mu}=(t,{\bf x})$, where
$t=x^0$ is the time coordinate and ${\bf x}=(x^1,x^2,x^3)$ represents the
space coordinates.

The physical objects that we study are particles living in spacetime. By a particle 
we mean a material point in space. 
More precisely, since the concept of space is not a well-defined entity 
in relativity, a better definition of a particle is a curve in spacetime. Thus, 
the particle is a 1-dimensional object living in the 4-dimensional spacetime.
 
The simplest way to specify a curve is through a set of 4 equations
\begin{equation}\label{nikolic:book1}
x^{\mu}=X^{\mu}(s) ,
\end{equation}
where $s$ is an auxiliary real parameter and $X^{\mu}(s)$ are some 
specified functions of $s$. Each $s$ defines one point on the curve and the set
of all values of $s$ defines the whole curve. In this sense, the curve can be identified
with the functions $X^{\mu}(s)$.

The parameter $s$ is a scalar with respect to Lorentz transformations or any other transformations
of the spacetime coordinates $x^{\mu}$. In this sense, the parametric definition
of the curve (\ref{nikolic:book1}) is covariant. However, non-covariant definitions are also possible.
For example, if the function $X^{0}(s)$ can be inverted, then the inverse $s(X^0)$
can be plugged into the space components  $X^{i}(s(X^0))\equiv \tilde{X}^{i}(X^0)$.
This leads to the usual nonrelativistic view of the particle as an object with the trajectory
$x^i=\tilde{X}^{i}(X^0)$, where $X^0$ is time.

{\it A priori}, the auxiliary parameter $s$ does not have any physical interpretation.
It is merely a mathematical parameter that cannot be measured. In fact, a transformation 
of the form
\begin{equation}\label{nikolic:book2}
 s\rightarrow s'=f(s) 
\end{equation}
does not change the curve in spacetime.\footnote{The only restriction on the function
$f(s)$ is that $df(s)/ds >0$.} This means that the functions $X^{\mu}(s)$
and $\tilde{X}^{\mu}(s)\equiv X^{\mu}(f(s))$ represent the same curve.

Since the curve is a 1-dimensional manifold, the parameter $s$ can be viewed as a
coordinate on that manifold. The transformation (\ref{nikolic:book2}) is a 
coodinate transformation on that manifold. One can also define the metric
tensor $h(s)$ on that manifold, such that $h(s)ds^2$ is the (squared) infinitesimal
length of the curve. Since the manifold is 1-dimensional, the metric tensor
$h$ has only 1 component. It is important to stress that this is an
intrinsic definition of the length of the curve that may be defined completely
independently on the spacetime metric $\eta_{\mu\nu}$.
This intrinsic length is not measurable so one can freely choose the metric $h(s)$.
However, once $h(s)$ is chosen, the metric in any other coordinate $s'$ is defined
through
\begin{equation}\label{nikolic:book3}
 h(s)ds^2 = h'(s')ds'^2 .
\end{equation}

We say that the curve at a point $s$ is timelike if the spacetime vector tangent to the curve at this
point is timelike. Spacelike and lightlike parts of the curve are defined analogously.
Thus, the part of the curve is timelike if $\dot{X}^{\mu} \dot{X}_{\mu}>0$,
spacelike if $\dot{X}^{\mu} \dot{X}_{\mu}<0$, and lightlike if $\dot{X}^{\mu} \dot{X}_{\mu}=0$, where $\dot{X}^{\mu}=dX^{\mu}(s)/ds$.\footnote{Here $A^{\mu} B_{\mu}\equiv \eta_{\mu\nu}A^{\mu} B^{\nu}$ and the summation
over repeated vector indices $\mu$, $\nu$ is understood.}
A timelike trajectory describes a particle that moves slower than light, a lightlike trajectory
describes a particle that moves with the velocity of light, and a spacelike
trajectory describes a particle that moves faster than light.
Contrary to what one might expect, we see that relativistic kinematics allows
particles to move even faster than light. As we shall see in the next subsection,
it is relativistic dynamics that may (or may not!) forbid motions faster than light,
depending on details of the dynamics.

For a timelike trajectory, there exists one special choice of the parameter $s$.
Namely, one can choose it to be equal to the proper time $\tau$ defined by
\begin{equation}\label{nikolic:book4}
 d\tau^2=dX^{\mu}dX_{\mu} .
\end{equation}
For such a choice, we see that
\begin{equation}\label{nikolic:book5}
\dot{X}^{\mu} \dot{X}_{\mu}=1 .
\end{equation}
In this case it is convenient to choose the metric on the trajectory such that
$h(\tau)=1$, so that the intrinsic length of the curve coincides with the
proper time, which, by definition, is equal to the extrinsic length defined by the
spacetime metric $\eta_{\mu\nu}$. Yet, such a choice is by no means
necessary.

Finally, let us briefly generalize the results above to the case of many particles.
If there are $n$ particles, then they are described by $n$ trajectories
$X_a^{\mu}(s_a)$, $a=1,\ldots,n$. Note that each trajectory is parameterized by its own
parameter $s_a$. However, since the parameterization of each curve is arbitrary,
one may parameterize all trajectories by the same parameter $s$, so that
the trajectories are described by the functions $X_a^{\mu}(s)$. 
In fact, the functions $X_a^{\mu}(s)$, which describe $n$ curves in the
4-dimensional spacetime, can also be viewed as {\em one} curve on a $4n$-dimensional
manifold with coordinates $x_a^{\mu}$. 

\subsection{Dynamics}
\label{nikolic:Dynamics}

\subsubsection{Action and equations of motion}

Dynamics of a relativistic particle is described by an action of the form
\begin{equation}\label{nikolic:book6}
 A=\int ds \, L(X(s),\dot{X}(s),s) ,
\end{equation}
where $X\equiv \{ X^{\mu} \}$, $\dot{X}\equiv \{ \dot{X}^{\mu} \}$.
We require that the Lagrangian $L$ should be a scalar with respect to spacetime coordinate
transformations. This means that all spacetime indices $\mu$ must be contracted.
We also require that the action should be invariant with respect to
reparameterizations of the form of (\ref{nikolic:book2}). From (\ref{nikolic:book3}), we see that this 
implies that any $ds$ should by multiplied by $\sqrt{h(s)}$, because such a product
is invariant with respect to (\ref{nikolic:book2}). To restrict the dependence on $s$
as much as possible, we assume that there is no other explicit dependence on $s$
except through the dependence on $h(s)$.
To further restrict the possible forms
of the action, we require that $L$ should be at most quadratic in the velocities 
$\dot{X}^{\mu}(s)$. With these requirements, the most general action can be written
in the form
\begin{equation}\label{nikolic:book7}
 A=-\int ds \, \sqrt{h(s)} \left[ \frac{1}{2h(s)}  
\frac{dX^\mu}{ds} \frac{dX^\nu}{ds} C_{\mu\nu}(X)
+\frac{1}{\sqrt{h(s)}} \frac{dX^\mu}{ds} C_{\mu}(X) + C(X) \right] .
\end{equation}
The functions $C(X)$, $C_{\mu}(X)$, and $C_{\mu\nu}(X)$ are referred to as
scalar potential, vector potential, and tensor potential, respectively.

What is the dynamical role of the function $h(s)$? Requiring that $h(s)$ is a
dynamical variable, the dynamical equation of motion $\delta A/\delta h(s)=0$
leads to
\begin{equation}\label{nikolic:book8}
 h^{-1}C_{\mu\nu}(X)\dot{X}^{\mu}\dot{X}^{\nu}=2C(X) .
\end{equation}
Viewed as an equation for $h$, it can be trivially solved as 
$h=C_{\mu\nu}\dot{X}^{\mu}\dot{X}^{\nu}/2C$. However,
since $h$ is not a physical quantity, this solution does not bring an important
physical information. Nevertheless, Eq.~(\ref{nikolic:book8}) does play an important
physical role, as we shall see soon.

Eq.~(\ref{nikolic:book8}) determines $h(s)$ only when the coordinate $s$ is chosen.
Thus, $h(s)$ can still be changed by changing the coordinate. In particular,
from (\ref{nikolic:book3}) we see that the coordinate transformation of the form
$s'(s)={\rm const}\int ds \, \sqrt{h(s)}$ makes $h'(s')$ a constant.
Thus, omitting the prime, we can fix $\sqrt{h(s)}=m^{-1}$, where $m$ is a positive constant.
For convenience, we choose $s$ to have the dimension of time and $C_{\mu\nu}$ 
to be dimensionless. Then the action (\ref{nikolic:book7}) implies that $m$ has the dimension
of mass (recall that we work in units $c=1$). Hence, we can rewrite (\ref{nikolic:book7})
as
\begin{equation}\label{nikolic:book9}
 A=-\int ds \left[ \frac{m}{2}  C_{\mu\nu}(X)
\dot{X}^{\mu} \dot{X}^{\nu} 
+C_{\mu}(X) \dot{X}^{\mu} + \frac{C(X)}{m} \right] .
\end{equation}
Now $m$ is no longer a dynamical quantity, but Eq.~(\ref{nikolic:book8}) rewritten as
\begin{equation}\label{nikolic:book10}
 C_{\mu\nu}(X)\dot{X}^{\mu}\dot{X}^{\nu}=\frac{2C(X)}{m^2} 
\end{equation}
should be added to (\ref{nikolic:book9}) as an additional constraint.

Now we are ready to study the physical role of the potentials $C$, $C_{\mu}$ and $C_{\mu\nu}$.
By writing $C_{\mu}(x)\equiv e A_{\mu}(x)$, one recognizes that the second term in
(\ref{nikolic:book9}) looks just like the action for the particle with the charge $e$ 
moving under the influence of the external electromagnetic potential $A_{\mu}(x)$
(see, e.g., \cite{jackson}). 
Similarly, by writing $C_{\mu\nu}(x)\equiv g_{\mu\nu}(x)$, one recognizes that the first term in
(\ref{nikolic:book9}) looks just like the action for the particle moving in a gravitational
background described by the curved metric tensor $g_{\mu\nu}(x)$ (see, e.g., \cite{weinberg}).
Since the physical properties of electromagnetic and gravitational forces are 
well known, we shall not study them in further discussions. Instead, from now on
we assume $C_{\mu}(x)=0$, $C_{\mu\nu}(x)=\eta_{\mu\nu}$.
Therefore, introducing the notation $U(X)\equiv C(X)/m$,
Eqs.~(\ref{nikolic:book9}) and (\ref{nikolic:book10}) reduce to
\begin{equation}\label{nikolic:book11}
 A=-\int ds \left[ \frac{m}{2} \dot{X}^{\mu} \dot{X}_{\mu} + U(X) \right]  ,
\end{equation}
\begin{equation}\label{nikolic:book12}
 \dot{X}^{\mu}\dot{X}_{\mu}=\frac{2U(X)}{m} .
\end{equation}
We see that the scalar potential $U(X)$ has the dimension of energy.
The dynamical equation of motion for $X^{\mu}(s)$ is 
$\delta A/\delta X^{\mu}(s)=0$. Applying this to (\ref{nikolic:book11}), one obtains
a relativistic Newton equation
\begin{equation}\label{nikolic:book13}
 m\frac{d^2X^{\mu}(s)}{ds^2}=\partial^{\mu}U(X(s)) ,
\end{equation}
where $\partial^{\mu}\equiv \eta^{\mu\nu}\partial/\partial X^{\nu}$.
The constraint (\ref{nikolic:book12}) is compatible with (\ref{nikolic:book13}).
Indeed, by applying the derivative $d/ds$ on (\ref{nikolic:book12}), one obtains
\begin{equation}
 [m\ddot{X}^{\mu}-\partial^{\mu}U(X)] \dot{X}_{\mu}=0 ,
\end{equation}
which is consistent because the expression in the square bracket trivially vanishes
when (\ref{nikolic:book13}) is satisfied.

The constraint (\ref{nikolic:book12}) implies
that the sign of $\dot{X}^{\mu}\dot{X}_{\mu}$ is equal to the sign of
$U$. Thus, we see that the particle moves slower than light if $U>0$,
with the velocity of light if $U=0$, and faster than light if $U<0$. 
Since $U(X)$ may change sign as $X$ varies, we see that the particle may, e.g., 
start motion with a velocity slower than light and {\em accelerate to a velocity
faster than light}.

At first sight, one may think that acceleration to velocities faster than light
is in contradiction with the well-known ``fact'' that the principle of relativity
does not allow particles to accelerate to velocities faster than light. 
However, there is no contradiction because this well-known ``fact'' is valid
only if some additional assumptions are fulfilled. In particular, if all forces
on particles are either of the electromagnetic type (vector potential) 
or of the gravitational type (tensor potential), then acceleration to velocities
faster than light is forbidden. Indeed, as far as we know, 
all relativistic classical forces on particles that exist in nature are of those two types.
Nevertheless, the principle of relativity allows also relativistic forces based
on the scalar potential, which, as we have seen, does allow acceleration to velocities
faster than light. Such classical forces have not yet been found in nature,
but it does not imply that they are forbidden. More precisely,
they may be forbidden by some additional physical principle taken together with the principle of relativity, but they 
are {\em not forbidden by the principle of relativity alone}.
 
\subsubsection{Canonical momentum and the Hamilton-Jacobi formulation}

Physics defined by (\ref{nikolic:book11})-(\ref{nikolic:book12}) can also be described by introducing
the canonical momentum
\begin{equation}\label{nikolic:book14}
 P_{\mu}=\frac{\partial L}{\partial \dot{X}^{\mu}} ,
\end{equation}
where  
\begin{equation}\label{nikolic:book15}
 L(X,\dot{X})=- \frac{m}{2} \dot{X}^{\mu} \dot{X}_{\mu} - U(X)   .
\end{equation}
This leads to
\begin{equation}\label{nikolic:book16}
 P^{\mu}=-m\dot{X}^{\mu} .
\end{equation}
The canonical Hamiltonian is
\begin{equation}\label{nikolic:book17}
H(P,X)=P_{\mu}\dot{X}^{\mu}-L=-\frac{P^{\mu}P_{\mu}}{2m}+U(X) .
\end{equation}
Note that this Hamiltonian is {\em not} the energy of the particle. In particular,
while particle energy transforms as a time-component of a spacetime vector,
the Hamiltonian above transforms as a scalar. This is a consequence of the fact 
$\dot{X}^{\mu}$ is not a derivative with respect to time $x^0$, but a derivative
with respect to the scalar $s$.

The constraint (\ref{nikolic:book12}) now can be written as
\begin{equation}\label{nikolic:book18}
 P^{\mu}P_{\mu}=2mU(X) .
\end{equation}
In relativity, it is customary to {\em define} the invariant mass $M$ through
the identity $P^{\mu}P_{\mu}\equiv M^2$. This shows that the mass depends
on $X$ as
\begin{equation}\label{nikolic:book19}
 M^2(X)=2mU(X) .
\end{equation}
Since $U(X)$ may change sign as $X$ varies, we see that the particle may, e.g., 
start motion as an ``ordinary'' massive particle ($M^2>0$) and later evolve into a
tachyon ($M^2<0$). The usual proof that an ``ordinary'' particle cannot reach
(or exceed) the velocity of light involves an assumption that the mass is a constant.
When mass is not a constant, or more precisely when $M^2$ can change sign,
then particle can reach and exceed the velocity of light.
 
The existence of the Hamiltonian allows us to formulate classical relativistic mechanics
with the relativistic Hamilton-Jacobi formalism.
One introduces the scalar Hamilton-Jacobi function $S(x,s)$ satisfying the 
Hamilton-Jacobi equation
\begin{equation}\label{nikolic:book21}
H(\partial S,x)=-\frac{\partial S}{\partial s} .
\end{equation}
Comparing (\ref{nikolic:book18}) with (\ref{nikolic:book17}), we see that the constraint
(\ref{nikolic:book18}) can be written as
 \begin{equation}\label{nikolic:book20}
  H(P,X)=0 .
 \end{equation}
The constraint (\ref{nikolic:book20}) implies that the right-hand side of (\ref{nikolic:book21})
must vanish, i.e., that $S(x,s)=S(x)$.
Hence (\ref{nikolic:book21}) reduces to $H(\partial S,x)=0$, which
in an explicit form reads
\begin{equation}\label{nikolic:book22}
 -\frac{(\partial^{\mu}S) (\partial_{\mu}S)}{2m}+U(x) =0 .
\end{equation}
The solution $S(x)$ determines the particle momentum 
\begin{equation}\label{nikolic:book23}
 P^{\mu}=\partial^{\mu}S(X) ,
\end{equation}
which, through (\ref{nikolic:book16}), determines the particle trajectory
\begin{equation}\label{nikolic:book24}
 \frac{dX^{\mu}(s)}{ds}=-\frac{\partial^{\mu}S(X(s))}{m} .
\end{equation}

\subsubsection{Generalization to many particles}

Now, let us briefly generalize all this to the case of many particles.
We study the dynamics of $n$ trajectories $X_a^{\mu}(s)$, $a=1,\ldots,n$, 
parameterized by a single parameter $s$.
In the general action (\ref{nikolic:book7}), the velocity-dependent terms
generalize as follows
\begin{equation}
\dot{X}^{\mu}C_{\mu} \rightarrow \sum_{a=1}^n \dot{X}_a^{\mu} C_{a\mu} ,
\end{equation}
\begin{equation}
\dot{X}^{\mu}\dot{X}^{\nu}C_{\mu\nu} \rightarrow 
\sum_{a=1}^n  \sum_{b=1}^n \dot{X}_a^{\mu} \dot{X}_b^{\nu}C_{ab\mu\nu} .
\end{equation}
Since the scalar potential is our main concern, we consider trivial vector and tensor potentials
$C_{a\mu}=0$ and $C_{ab\mu\nu}=c_a\delta_{ab}\eta_{\mu\nu}$, respectively, where
$c_a$ are constants.
Thus, Eqs.~(\ref{nikolic:book11})-(\ref{nikolic:book12}) generalize to
\begin{equation}\label{nikolic:book25}
 A=-\int ds \left[  \sum_{a=1}^n \frac{m_a}{2}
\dot{X}_a^{\mu} \dot{X}_{a\mu} + 
U(X_1,\ldots,X_n) \right]  ,
\end{equation}
\begin{equation}\label{nikolic:book26}
 \sum_{a=1}^n m_a \dot{X}_a^{\mu}\dot{X}_{a\mu}=2U(X_1,\ldots,X_n) ,
\end{equation}
where $c_a$ are dimensionless and $m_a=mc_a$.
The relativistic Newton equation (\ref{nikolic:book13}) generalizes to
\begin{equation}\label{nikolic:book27}
 m_a\frac{d^2X_a^{\mu}(s)}{ds^2}=\partial_a^{\mu}U(X_1(s),\ldots,X_n(s)) .
\end{equation}
In general, from (\ref{nikolic:book27}) we see that the force on the particle $a$
at the spacetime position $X_a(s)$ depends on positions of all other particles
for the same $s$. In other words, the forces on particles are nonlocal. Nevertheless,
since $s$ is a scalar, such nonlocal forces are compatible with the principle of relativity;
the nonlocal equation of motion (\ref{nikolic:book27}) is relativistic covariant.
Thus we see that {\em relativity and nonlocality are compatible with each other}.
Even though for each $s$ there may exist a particular ($s$-dependent) 
Lorentz frame with respect to which
the force between two particles is instantaneous, such a Lorentz frame is by no means
special or ``preferred''. Instead, such a particular Lorentz frame is determined by covariant
equations of motion supplemented by a particular choice of initial
conditions $X_a^{\mu}(0)$.
(Of course, the initial velocities $\dot{X}_a^{\mu}(0)$ also need to be 
chosen for a solution of (\ref{nikolic:book27}), but the initial
velocities can be specified in a covariant manner through the equation
(\ref{nikolic:book31}) below.)

Note also that the phenomena of {\em nonlocal forces
between particles} and {\em particle motions faster than light} are independent of each other.
The force (\ref{nikolic:book27}) becomes local when
\begin{equation}\label{nikolic:book28}
 U(X_1,\ldots,X_n)=U_1(X_1)+ \cdots + U_n(X_n) ,
\end{equation}
in which case (\ref{nikolic:book27}) reduces to
\begin{equation}\label{nikolic:book29}
 m_a\frac{d^2X_a^{\mu}(s)}{ds^2}=\partial_a^{\mu}U_a(X_a(s)) .
\end{equation}
Thus we see that particle motions faster than light ($U_a<0$)
are possible even when the forces are local.
Similarly, $U(X_1,\ldots,X_n)$ may be such that particles move only slower than light,
but that the forces are still nonlocal. 

The Hamilton-Jacobi formalism can also be generalized to the many-particle case.
In particular, Eqs.~(\ref{nikolic:book22}) and (\ref{nikolic:book24}) generalize to
\begin{equation}\label{nikolic:book30}
 -\sum_{a=1}^n \frac{(\partial_a^{\mu}S) (\partial_{a\mu}S)}{2m_a}
+U(x_1,\ldots, x_n) =0 ,
\end{equation}
\begin{equation}\label{nikolic:book31}
 \frac{dX_a^{\mu}(s)}{ds}=-\frac{\partial_a^{\mu}S(X_1(s),\ldots, X_n(s))}{m_a} ,
\end{equation}
respectively.
In the local case (\ref{nikolic:book28}), the solution of (\ref{nikolic:book30}) can be written in the form
\begin{equation}\label{nikolic:book32}
 S(x_1,\ldots ,x_n)=S_1(x_1)+ \cdots + S_n(x_n) ,
\end{equation}
so (\ref{nikolic:book31}) reduces to
\begin{equation}\label{nikolic:book33}
 \frac{dX_a^{\mu}(s)}{ds}=-\frac{\partial_a^{\mu}S_a(X_a(s))}{m_a} .
\end{equation}

\subsubsection{Absolute time}

Finally, let us give a few conceptual remarks on the physical meaning of the parameter $s$.
As discussed in more detail in \cite{nikscalpot,niktimeprob}, its role in the equations above is 
formally analogous to the role of the Newton absolute time $t$ in nonrelativistic 
Newtonian mechanics. In particular, even though $s$ cannot be measured directly,
it can be measured indirectly in the same sense as $t$ is measured indirectly in Newtonian mechanics.
Namely, one measures time by a ``clock'', where ``clock'' is nothing but a physical process 
periodic in time. Hence, if at least one of the $4n$ functions
$X_a^{\mu}(s)$ is periodic in $s$, then the number of cycles
(which is a measurable quantity) can be interpreted as a measure of elapsed $s$.
Thus, it is justified to think of $s$ as an absolute time in relativistic mechanics.

The parameter $s$ is also related to the more familiar relativistic notion of proper time $\tau$.
As discussed in more detail in \cite{nikscalpot,niktimeprob}, $s$ can be thought of as a generalization
of the notion of proper time.

\section{Relativistic quantum mechanics}
\label{nikolic:RQ}

\subsection{Wave functions and their relativistic probabilistic interpretation}
\label{nikolic:RQ1}

Let us start with quantum mechanics of a single particle without spin.
The basic object describing the properties of the particle is the wave function
$\psi(x)$. We normalize the wave function such that
\begin{equation}\label{nikolic:book34}
\int d^4x \, \psi^*(x)\psi(x)=1 .
\end{equation}
More precisely, to avoid a divergence,
the integral $\int d^4x$ is taken over some very large but not necessarily
infinite 4-dimensional region. (For most practical purposes it is 
more than sufficient
to take a region of the astronomical size.) If the integral (\ref{nikolic:book34})
happens to converge
even when the boundary of the region is at infinity, then an infinite 4-dimensional
region is also allowed.

The probability of finding the particle in the (infinitesimal) 4-volume $d^4x$
is postulated to be
\begin{equation}\label{nikolic:e6}
dP=|\psi(x)|^2d^4x ,
\end{equation}
which is compatible with the normalization (\ref{nikolic:book34}), as $|\psi|^2 \equiv \psi^*\psi$.
At first sight, (\ref{nikolic:e6}) may seem to be incompatible with the usual
probabilistic interpretation in 3-space\footnote{To our knowledge, the first version of probabilistic interpretation
based on (\ref{nikolic:e6}) rather than (\ref{nikolic:e7}) was proposed in \cite{stuc}.} 
\begin{equation}\label{nikolic:e7}
dP_{(3)}\propto |\psi({\bf x},t)|^2d^3x .
\end{equation}
Nevertheless, (\ref{nikolic:e6}) is compatible with (\ref{nikolic:e7}). If (\ref{nikolic:e6})
is the fundamental {\it a priori} probability, then (\ref{nikolic:e7})
is naturally interpreted as the conditional probability corresponding to the case
in which one knows that the particle is detected at time $t$.
More precisely, the conditional probability is
\begin{equation}\label{nikolic:e8}
dP_{(3)}=\frac{|\psi({\bf x},t)|^2 d^3x}{N_t},
\end{equation}
where 
\begin{equation}\label{nikolic:e9}
 N_t=\int d^3x |\psi({\bf x},t)|^2 
\end{equation}
is the normalization factor. If $\psi$ is normalized such that (\ref{nikolic:e6}) is valid,
then (\ref{nikolic:e9}) is also the marginal probability that the particle will be found
at $t$. Of course, in practice a measurement always lasts a finite time $\Delta t$
and the detection time $t$ cannot be determined with perfect accuracy. 
Thus, (\ref{nikolic:e8}) should be viewed as a limiting case in which
the fundamental probability (\ref{nikolic:e6}) is averaged over a very small $\Delta t$.
More precisely, if the particle is detected 
between $t-\Delta t/2$ and $t+\Delta t/2$, then  
(\ref{nikolic:e8}) is the probability of different 3-space positions of the particle detected 
during this small $\Delta t$.

Can the probabilistic interpretation (\ref{nikolic:e6}) be verified experimentally?
In fact, it already is! In practice one often measures cross sections 
associated with scattering experiments or decay widths and lifetimes
associated with spontaneous decays of unstable quantum systems. 
These experiments agree with standard theoretical predictions. Our point is
that these standard theoretical predictions actually use (\ref{nikolic:e6}),
although not explicitly. Let us briefly explain it. 
The basic theoretical tool
in these predictions is the transition amplitude $A$. Essentially,
the transition amplitude is the wave function (usually Fourier transformed
to the 3-momentum space) at $t\rightarrow\infty$, calculated by 
assuming that the wave function at $t\rightarrow -\infty$ is known.
Due to energy conservation one obtains 
\begin{equation}
 A\propto\delta(E_{\rm in}-E_{\rm fin}) ,
\end{equation}
where $E_{\rm in}$ and $E_{\rm fin}$ are the initial and final energy, respectively.
Thus, the transition probability is proportional to
\begin{equation}
|A|^2\propto[\delta(E_{\rm in}-E_{\rm fin})]^2=\frac{T}{2\pi}\delta(E_{\rm in}-E_{\rm fin}),
\end{equation}
where $T=\int dt =2\pi \delta(E=0)$ and we work in units $\hbar=1$. 
Since $T$ is infinite, this transition
probability is physically meaningless. The standard interpretation 
(see, e.g., \cite{schiff} for the nonrelativistic case or \cite{halz,bd1} for the
relativistic case), which agrees with experiments, 
is that the physical quantity is $|A|^2/T$ and that this quantity is (proportional to)
the transition probability {\em per unit time}. But this is essentially the same
as our equation (\ref{nikolic:e6}) which says that $\int d^3x |\psi|^2$ is not 
probability itself, but probability {\em per unit time}. Although 
the interpretation of $|A|^2/T$ as probability per unit time 
may seem plausible even without explicitly postulating (\ref{nikolic:e6}),
without this postulate such an interpretation of $|A|^2/T$
is at best heuristic and cannot be strictly derived from other basic
postulates of QM, including (\ref{nikolic:e7}).
In this sense, the standard interpretation of transition amplitudes
in terms of transition probabilities per unit time
is better founded in basic axioms of QM if (\ref{nikolic:e6}) is also adopted
as one of its axioms.

Now let us generalize it to the case of $n$ particles.
Each particle has its own space 
position ${\bf x}_a$, $a=1,\ldots,n$, as well as its own time coordinate
$t_a$. Therefore, the wave function is of the form $\psi(x_1,\ldots,x_n)$,
which is a many-time wave function. (For an early use of many-time
wave functions in QM see \cite{tomon}). 
Then (\ref{nikolic:e6}) generalizes to
\begin{equation}\label{nikolic:e6n}
dP=|\psi(x_1,\ldots,x_n)|^2 d^4x_1 \cdots d^4x_n .
\end{equation}
Hence, if the first particle is detected at $t_1$, second particle at $t_2$, etc., 
then Eq.~(\ref{nikolic:e8}) generalizes to
\begin{equation}\label{nikolic:e7n}
dP_{(3n)}=\frac{|\psi({\bf x}_1,t_1,\ldots,{\bf x}_n,t_n)|^2 d^3x_1 \cdots d^3x_n}
{N_{t_1,\ldots,t_n}} ,
\end{equation}
where 
\begin{equation}\label{nikolic:e8n}
N_{t_1,\ldots,t_n}=\int |\psi({\bf x}_1,t_1,\ldots,{\bf x}_n,t_n)|^2 d^3x_1 \cdots d^3x_n .
\end{equation}

The many-time wave function contains also the
familiar single-time wave function as a special case
\begin{equation}\label{nikolic:stime}
 \psi({\bf x}_1,\ldots,{\bf x}_n;t)=\psi({\bf x}_1,t_1,\ldots,{\bf x}_n,t_n) 
|_{t_1=\cdots=t_n=t} .
\end{equation}
In this case  (\ref{nikolic:e7n}) reduces to the familiar expression
\begin{equation}\label{nikolic:e7ns}
dP_{(3n)}=\frac{|\psi({\bf x}_1,\ldots,{\bf x}_n;t)|^2 d^3x_1 \cdots d^3x_n}
{N_{t}} ,
\end{equation}
where $N_{t}$ is given by (\ref{nikolic:e8n}) calculated at $t_1=\cdots=t_n=t$.

Finally, let us generalize all this to particles that carry spin or some other additional
discrete degree of freedom. For one particle, instead of one wave function
$\psi(x)$, one deals with a collection of wave functions $\psi_l(x)$, where $l$ is a discrete label.
Similarly, for $n$ particles with discrete degrees of freedom we have 
a collection of wave functions of the form $\psi_{l_1\dots l_n}(x_1,\ldots,x_n)$.
To simplify the notation, it is convenient to introduce a 
collective label $L=(l_1,\ldots,l_n)$, which means that the wave function for
$n$ particles can be written as $\psi_{L}(x_1,\ldots,x_n)$.
Now all equations above can be easily generalized through the replacement
\begin{equation}
 \psi^*\psi \rightarrow \sum_L \psi_L^*\psi_L . 
\end{equation}
In particular, the joint probability for finding particles at the positions
$x_1,\ldots,x_n$ is given by a generalization of (\ref{nikolic:e6n}) 
\begin{equation}\label{nikolic:e6ng}
dP=\sum_L \psi_L^*(x_1,\ldots,x_n) \psi_L(x_1,\ldots,x_n) d^4x_1 \cdots d^4x_n .
\end{equation}
Another useful notation is to introduce the column $\psi=\{ \psi_L \}$ 
and the row $\psi^{\dagger}=\{ \psi^*_L \}$, i.e., 
\begin{equation}
\psi=\left( 
\begin{array}{c} 
 \psi_1 \\ \psi_2 \\ \vdots
\end{array}
\right) , \;\;\;\;
 \psi^{\dagger}=\left( 
\begin{array}{ccc} 
 \psi_1^* & \psi_2^* & \cdots
\end{array}
\right) .
\end{equation}
With this notation, (\ref{nikolic:e6ng}) can also be written as
\begin{equation}\label{nikolic:e6ng2}
dP=\psi^{\dagger}(x_1,\ldots,x_n) \psi(x_1,\ldots,x_n) d^4x_1 \cdots d^4x_n .
\end{equation}

\subsection{Theory of quantum measurements}
\label{nikolic:RQ2}

Let $\psi(x)$ be expanded as
\begin{equation}\label{nikolic:meas0}
 \psi(x)=\sum_b c_b \psi_b(x) ,
\end{equation}
where $\psi_b(x)$ are eigenstates of some hermitian operator $\hat{B}$ on the Hilbert space
of functions of $x$. Let $\psi_b(x)$ be normalized such that 
$\int d^4x \, \psi^*_b(x)\psi_b(x) =1$. Assume that one measures the value
of the observable $B$ described by the hermitian operator $\hat{B}$.
In a conventional approach to QM, one would postulate that $|c_b|^2$ is the probability that 
$B$ will take the value $b$. Nevertheless, there is no need for such a postulate
because, whatever the operator $\hat{B}$ is, this probabilistic rule can be derived
from the probabilistic interpretation in the position space discussed in Sec.~\ref{nikolic:RQ1}. 

To understand this, one needs to understand how a typical measuring apparatus works,
i.e., how the wave function of the measured system described by the coordinate $x$
interacts with the wave function of the measuring apparatus described by the coordinate $y$.
(For simplicity, we assume that $y$ is a coordinate of a single particle, but essentially
the same analysis can be given by considering a more realistic case in which
$y$ is replaced by a macroscopically large number $N$ of particles
$y_1,\ldots,y_N$ describing the macroscopic measuring apparatus. Similarly,
the same analysis can also be generalized to the case in which $x$ is replaced by
$x_1,\ldots,x_n$.)
Let the wave function of the measuring apparatus for times before the interaction
be $E_0(y)$. Thus, for times $x^0$ and $y^0$ before the interaction, the total wave function is
$\psi(x)E_0(y)$. But what happens after the interaction?
If $\psi(x)=\psi_b(x)$ before the interaction, then the interaction must be such that
after the interaction the total wave function takes the form $\psi_b(x)E_b(y)$, 
where $E_b(y)$ is a macroscopic state 
of the measuring apparatus, normalized so that $\int d^4y \, E^*_b(y) E_b(y) =1$.
The state $E_b(y)$ is such that
one can say that ``the measuring apparatus shows that the result of measurement is $b$''
when the measuring apparatus is found in that state.
Schematically, the result of interaction described above can be written as
\begin{equation}\label{nikolic:meas1}
\psi_b(x) E_0(y) \rightarrow \psi_b(x)E_b(y) .
\end{equation}
Of course, most interactions do not have the form (\ref{nikolic:meas1}), but only those
that do can be regarded as measurements of the observable $\hat{B}$.
The transition (\ref{nikolic:meas1}) is guided by some {\em linear} differential equation
(we study the explicit linear dynamical equations for wave functions in the subsequent sections),
which means that the superposition principle is valid.
Therefore, (\ref{nikolic:meas1}) implies that for a general superposition (\ref{nikolic:meas0})
we have
\begin{equation}\label{nikolic:meas2}
\sum_b c_b \psi_b(x) E_0(y) \rightarrow \sum_b c_b \psi_b(x)E_b(y) \equiv \psi(x,y).
\end{equation}

The states $E_b(y)$ must be macroscopically distinguishable. In practice, it means
that they do not overlap (or more realistically that their overlap is negligible), i.e., that
\begin{equation}\label{nikolic:meas3}
E_b(y) E_{b'}(y) \simeq 0 \;\; {\rm for} \;\; b \neq b' ,
\end{equation}
for all values of $y$. Instead of asking ``what is the probability that the measured particle
is in the state $\psi_b(x)$'', the operationally more meaningfull question is
``what is the probability that the measuring apparatus will be found
in the state $E_b(y)$''. The (marginal) probability density for finding the particle describing 
the measuring apparatus at the position $y$ is
\begin{equation}\label{nikolic:meas4}
 \rho(y)=\int d^4x \, \psi^*(x,y)\psi(x,y) .
\end{equation}
Using (\ref{nikolic:meas2}) and (\ref{nikolic:meas3}), this becomes
\begin{equation}\label{nikolic:meas5}
 \rho(y) \simeq \sum_b |c_b|^2 |E_b(y)|^2 .
\end{equation}
Now let ${\rm supp}\,E_b$ be the support of $E_b(y)$, i.e., the region of 
$y$-space on which $E_b(y)$ is not negligible. 
Then, from (\ref{nikolic:meas5}), the probability that $y$ will take a value from the
support of $E_b(y)$ is
\begin{equation}\label{nikolic:meas6}
 p_b=\int_{{\rm supp}\,E_b} d^4y \, \rho(y) \simeq  |c_b|^2 .
\end{equation}
In other words, the probability that the measuring apparatus will be found
in the state $E_b(y)$ is (approximately) equal to $|c_b|^2$.

\subsection{Relativistic wave equations}

In this subsection we consider particles which are free on the classical level,
i.e., particles classically described by the action (\ref{nikolic:book11})
with a constant scalar potential
\begin{equation}\label{nikolic:u=m/2}
 U(X)=\frac{m}{2} .
\end{equation}
The constraint (\ref{nikolic:book18}) becomes
\begin{equation}\label{nikolic:consc}
 P^{\mu}P_{\mu}-m^2 =0 ,
\end{equation}
implying that $m$ is the mass of the particle.

In QM, the momentum $P_{\mu}$ becomes the operator $\hat{P}_{\mu}$
satisfying the canonical commutation relations
\begin{equation}\label{nikolic:e3}
[x^{\mu},\hat{P}_{\nu}]=-i\eta^{\mu}_{\nu} ,
\end{equation}
where we work in units $\hbar=1$. These commutation relations are satisfied
by taking
\begin{equation}\label{nikolic:e2}
\hat{P}_{\nu}=i\partial_{\nu} .
\end{equation}

\subsubsection{Single particle without spin}

Let us start with a particle without spin. 
The quantum analog of the classical constraint (\ref{nikolic:consc}) is
\begin{equation}\label{nikolic:consq}
 [\hat{P}^{\mu}\hat{P}_{\mu}-m^2]\psi(x) =0 ,
\end{equation}
which is nothing but the Klein-Gordon equation
\begin{equation}\label{nikolic:KG}
 [\partial^{\mu}\partial_{\mu}+m^2]\psi(x) =0 .
\end{equation}
From a solution of  (\ref{nikolic:KG}), one can construct the real current
\begin{equation}\label{nikolic:loc2}
j_{\mu}=\frac{i}{2}\psi^*  \!\stackrel{\leftrightarrow\;}{\partial_{\mu}}\!   \psi ,
\end{equation}
where 
\begin{equation}
\psi_1  \!\stackrel{\leftrightarrow\;}{\partial_{\mu}}\!   \psi_2
\equiv  \psi_1 (\partial_{\mu}\psi_2) -  (\partial_{\mu}\psi_1) \psi_2 .
\end{equation}
Using (\ref{nikolic:KG}), one can show that this current is conserved
\begin{equation}\label{nikolic:conserv}
\partial_{\mu}j^{\mu}=0.
\end{equation}
By writing $\psi=Re^{iS}$, where $R$ and $S$ are real functions,
the complex Klein-Gordon equation (\ref{nikolic:KG}) is equivalent to a set of
two real equations
\begin{equation}\label{nikolic:cont}
\partial^{\mu}(R^2\partial_{\mu}S)=0,
\end{equation}
\begin{equation}\label{nikolic:HJ}
-\frac{(\partial^{\mu}S)(\partial_{\mu}S)}{2m} +\frac{m}{2} +Q=0,
\end{equation}
where (\ref{nikolic:cont}) is the conservation equation
(\ref{nikolic:conserv})  and
\begin{equation}\label{nikolic:Q=}
Q=\frac{1}{2m}\frac{\partial^{\mu}\partial_{\mu}R}{R} .
\end{equation}

It is easy to show that the equations above have the correct
nonrelativistic limit. In particular, by writing
\begin{equation}\label{nikolic:nr1}
\psi=\frac{e^{-imt}}{\sqrt{m}}\psi_{\rm NR}
\end{equation}
and using $|\partial_t\psi_{\rm NR}|\ll m|\psi_{\rm NR}|$,
$|\partial^2_t\psi_{\rm NR}|\ll m|\partial_t\psi_{\rm NR}|$,
from (\ref{nikolic:loc2}) and (\ref{nikolic:KG}) we find the approximate
equations
\begin{equation}
j_0=\psi_{\rm NR}^*\psi_{\rm NR},
\end{equation}
\begin{equation}\label{nikolic:nonrel}
-\frac{\nabla^2}{2m}\psi_{\rm NR}=i\partial_t\psi_{\rm NR},
\end{equation}
which are the nonrelativistic 
probability density and the 
nonrelativistic Schr\"o\-din\-ger equation for the evolution of the wave function
$\psi_{\rm NR}$, respectively.

Note that (\ref{nikolic:nr1}) contains a positive-frequency oscillatory function
$e^{-imt}$ and not a negative-frequency oscillatory function  $e^{imt}$.
If we took $e^{imt}$ in (\ref{nikolic:nr1}) instead, then we would obtain $-i\partial_t\psi_{\rm NR}$
on the right-hand side of (\ref{nikolic:nonrel}), which would be a Schr\"odinger equation
with the wrong sign of the time derivative. In other words, even though 
(\ref{nikolic:KG}) contains solutions with both positive and negative frequencies,
only positive frequencies lead to the correct nonrelativistic limit. This means that
only solutions with positive frequencies are physical, i.e., that the most general
physical solution of (\ref{nikolic:KG}) is
\begin{equation}\label{nikolic:e3.9f}
\psi(x)=\int d^3k \,  a({\bf k})
e^{-i[\omega({\bf k})x^0-{\bf k}{\bf x}]} , 
\end{equation}
where $a({\bf k})$ is an arbitrary function and
\begin{equation}\label{nikolic:e3.9'f}
\omega({\bf k})=\sqrt{{\bf k}^2+m^2}
\end{equation} 
is positive. More precisely, this is so if the particle is not charged, i.e., if the particle is its own antiparticle. When particles are charged, then $\psi$ with positive frequencies
describes a particle, while $\psi$ with negative frequencies
describes an antiparticle.

\subsubsection{Many particles without spin}

Now let us generalize it to the case of $n$ identical particles without spin,
with equal masses $m_a=m$.
The wave function $\psi$ satisfies $n$ Klein-Gordon equations
\begin{equation}\label{nikolic:KGn}
(\partial_a^{\mu}\partial_{a\mu}+m^2)\psi(x_1,\ldots ,x_n)=0 ,
\end{equation}
one for each $x_a$.
Therefore, one can introduce $n$ real 4-currents
\begin{equation}\label{nikolic:curn}
j^{\mu}_a=\frac{i}{2}\psi^* \!\stackrel{\leftrightarrow\;}{\partial^{\mu}_a}\! \psi ,
\end{equation}
each of which is separately conserved
\begin{equation}\label{nikolic:cons}
\partial^{\mu}_a j_{a\mu}=0.
\end{equation}
Equation (\ref{nikolic:KGn}) also implies
\begin{equation}\label{nikolic:KGs}
\left( \sum_a\partial_a^{\mu}\partial_{a\mu}+nm^2 \right)
\psi(x_1,\ldots ,x_n)=0 ,
\end{equation}
while (\ref{nikolic:cons}) implies
\begin{equation}\label{nikolic:conss}
\sum_a\partial^{\mu}_a j_{a\mu}=0.
\end{equation}
Next we write $\psi=Re^{iS}$, where $R$ and $S$ are real
functions. Equation (\ref{nikolic:KGs})
is then equivalent to a set of two real equations
\begin{equation}\label{nikolic:contn}
\sum_a\partial_a^{\mu}(R^2\partial_{a\mu}S)=0,
\end{equation}
\begin{equation}\label{nikolic:HJn}
-\frac{\sum_a(\partial_a^{\mu}S)(\partial_{a\mu}S)}{2m} +\frac{nm}{2} +Q=0,
\end{equation}
where
\begin{equation}\label{nikolic:Q}
Q=\frac{1}{2m}\frac{\sum_a\partial_a^{\mu}\partial_{a\mu}R}{R} .
\end{equation}
Eq.~(\ref{nikolic:contn}) is equivalent to (\ref{nikolic:conss}).

In the nonrelativistic limit we have $n$ equations of the form of (\ref{nikolic:nonrel})
\begin{equation}\label{nikolic:nonreln}
-\frac{\nabla_a^2}{2m}\psi_{\rm NR}=i\partial_{t_a}\psi_{\rm NR},
\end{equation}
where $\psi_{\rm NR}=\psi_{\rm NR}({\bf x}_1,t_1,\ldots,{\bf x}_n,t_n)$
is the nonrelativistic many-time wave function.
The single-time wave function is defined as in (\ref{nikolic:stime}), so we see that
\begin{equation}
\sum_a \partial_{t_a}\psi_{\rm NR}({\bf x}_1,t_1,\ldots,{\bf x}_n,t_n) 
|_{t_1=\cdots=t_n=t}  = \partial_{t} \psi_{\rm NR}({\bf x}_1,\ldots,{\bf x}_n;t) .
\end{equation}
Therefore (\ref{nikolic:nonreln}) implies the usual many-particle single-time Schr\"odinger
equation
\begin{equation}\label{nikolic:nonrelns}
\left[ \sum_a -\frac{\nabla_a^2}{2m} \right] \psi_{\rm NR}({\bf x}_1,\ldots,{\bf x}_n;t)
=i\partial_{t}\psi_{\rm NR}({\bf x}_1,\ldots,{\bf x}_n;t) .
\end{equation}

\subsubsection{Single particle with spin $\frac{1}{2}$}

A relativistic particle with spin $\frac{1}{2}$ is described by a 4-component wave function
$\psi_l(x)$, $l=1,2,3,4$ (see, e.g., \cite{bd1}). Each component satisfies
the Klein-Gordon equation
\begin{equation}\label{nikolic:KGl}
 [\partial^{\mu}\partial_{\mu}+m^2]\psi_l(x) =0 .
\end{equation} 
Introducing the column
\begin{equation} 
\psi=\left( 
\begin{array}{c} 
 \psi_1 \\ \psi_2 \\ \psi_3 \\ \psi_4
\end{array}
\right) ,
\end{equation}
known as Dirac spinor, (\ref{nikolic:KGl}) can also be written as
\begin{equation}\label{nikolic:KGsp}
 [\partial^{\mu}\partial_{\mu}+m^2]\psi(x) =0 .
\end{equation}
However, the 4 components of (\ref{nikolic:KGsp}) are not completely independent.
They also satisfy an additional constraint linear in the spacetime derivatives,
known as the Dirac equation
\begin{equation}\label{nikolic:dirac}
[i\gamma^{\bar{\mu}}\partial_{\mu}-m]\psi(x)=0 .
\end{equation}
Here each $\gamma^{\bar{\mu}}$ is a $4\times4$ matrix in the spinor space.
These matrices satisfy the anticommutation relations
\begin{equation}\label{nikolic:gamma}
 \gamma^{\bar{\mu}}\gamma^{\bar{\nu}}+\gamma^{\bar{\nu}}\gamma^{\bar{\mu}}
=2\eta^{\bar{\mu}\bar{\nu}} .
\end{equation}
In fact, by multiplying (\ref{nikolic:dirac}) from the left with the operator
$[-i\gamma^{\bar{\mu}}\partial_{\mu}-m]$ and using (\ref{nikolic:gamma}), one
obtains (\ref{nikolic:KGsp}). This means that the Klein-Gordon equation (\ref{nikolic:KGsp})
is a consequence of the Dirac equation (\ref{nikolic:dirac}). Note, however, that the
opposite is not true; one cannot derive (\ref{nikolic:dirac}) from (\ref{nikolic:KGsp}).

The matrices $\gamma^{\bar{\mu}}$ are known as Dirac matrices.
Even though they carry the index $\bar{\mu}$, they do {\em not} transform as
vectors under spacetime transformations. In fact, this is why $\bar{\mu}$ 
has a bar over it, to remind us that it is not a spacetime vector index.\footnote{
In most literature, like \cite{bd1}, the bar is omitted and the 
Dirac matrices are denoted by $\gamma^{\mu}$. In our opinion, such a notation
without a bar causes a lot of confusion.}
Instead, $\bar{\mu}$ is only a label.
More precisely, since $\gamma^{\bar{\mu}}$ does not carry any spacetime
index like $\mu$, it is a scalar with respect to spacetime transformations.
Similarly, the spinor $\psi$ also does not carry spacetime indices, so it is also
a scalar with respect to spacetime transformations.\footnote{
In most literature, like \cite{bd1}, the spinor $\psi$ transforms in a rather
complicated and unintuitive way under Lorentz transformations of spacetime
coordinates. Even worse, it turns out that such a complicated transformation of spinors
cannot be generalized to arbitrary transformations of spacetime coordinates.
This is why it is more convenient to adopt a more intuitive formalism in which
$\psi$ is a scalar with respect to spacetime transformations \cite{weinberg,birdel}.
Nevertheless, as long as only Lorentz transformations of physically measurable
quantities are concerned, the two formalisms turn out to be physically equivalent.}

Nevertheless, there is a way to introduce a matrix $\gamma^{\mu}$ that transforms 
as a true vector \cite{weinberg,birdel}.
At each point of spacetime, one introduces the tetrad $e^{\mu}_{\bar{\alpha}}(x)$, which
is a collection of four spacetime vectors, one for each $\bar{\alpha}=0,1,2,3$. 
The tetrad is chosen so that
\begin{equation}\label{nikolic:a1}
\eta^{\bar{\alpha}\bar{\beta}}e^{\mu}_{\bar{\alpha}}(x)e^{\nu}_{\bar{\beta}}(x)=
g^{\mu\nu}(x) ,
\end{equation}
where $g^{\mu\nu}(x)$ is the spacetime metric (which, in general, may depend on $x$)
and $\eta_{\bar{\alpha}\bar{\beta}}$ are 
components of a matrix equal to the Minkowski metric. 
The spacetime-vector indices are raised and lowered by $g^{\mu\nu}(x)$ and $g_{\mu\nu}(x)$,
respectively, while $\bar{\alpha}$-labels are raised and lowered by $\eta^{\bar{\alpha}\bar{\beta}}$ 
and $\eta_{\bar{\alpha}\bar{\beta}}$, respectively. Thus, (\ref{nikolic:a1}) can also be inverted as
\begin{equation}\label{nikolic:a2}
g^{\mu\nu}(x)e_{\mu}^{\bar{\alpha}}(x)e_{\nu}^{\bar{\beta}}(x)=
\eta^{\bar{\alpha}\bar{\beta}} .
\end{equation} 

Now from the constant Dirac matrices $\gamma^{\bar{\alpha}}$ we define
\begin{equation}\label{nikolic:a3}
\gamma^{\mu}(x)=e^{\mu}_{\bar{\alpha}}(x) \gamma^{\bar{\alpha}} .
\end{equation}   
The spinor indices carried by matrices $\gamma^{\bar{\alpha}}$ and $\gamma^{\mu}(x)$
are interpreted as indices of the spinor representation of the {\em internal} 
group SO(1,3). 
Just like $\psi(x)$, $\psi^{\dagger}(x)$ is also a scalar with respect to
spacetime coordinate transformations. It is also convenient to define the quantity
\begin{equation}\label{nikolic:a4}
\bar{\psi}(x)=\psi^{\dagger}(x) \gamma^{\bar{0}} ,
\end{equation}
which is also a scalar with respect to spacetime coordinate transformations.
Thus we see that the quantities 
\begin{equation}\label{nikolic:a5}
\bar{\psi}(x)\psi(x) , \;\;\;\; \psi^{\dagger}(x)\psi(x) ,
\end{equation}
are both scalars with respect to spacetime coordinate transformations, and 
that the quantities 
\begin{equation}\label{nikolic:a6}
\bar{\psi}(x)\gamma^{\mu}(x)\psi(x) , \;\;\;\; 
\frac{i}{2}\psi^{\dagger}(x)  \!\stackrel{\leftrightarrow\;}{\partial^{\mu}}\!  \psi(x) ,
\end{equation}
are both vectors with respect to spacetime coordinate transformations.

Note that in the flat Minkowski spacetime,
there is a particular global Lorentz frame of coordinates in which
\begin{equation}\label{nikolic:a7}
\gamma^{\mu}(x)=\gamma^{\bar{\mu}} .
\end{equation} 
Indeed, this is why Eq.~(\ref{nikolic:dirac}) makes sense. 
However, (\ref{nikolic:a7}) is not a covariant expression, but is only valid in one
special system of coordinates. 
In other global Lorentz frames we have
\begin{equation}
 \gamma'^{\mu}=\Lambda^{\mu}_{\;\;\nu}\gamma^{\nu} ,
\end{equation}
where $\Lambda^{\mu}_{\;\;\nu}$ are the matrix elements of the Lorentz transformation.
Since $\Lambda^{\mu}_{\;\;\nu}$ do not depend on $x$, it follows that 
the vector $\gamma^{\mu}$ is $x$-independent in {\em any} Lorentz frame.
Therefore, in an arbitrary Lorenz frame, (\ref{nikolic:dirac})
should be replaced by a truly Lorentz-covariant equation
\begin{equation}\label{nikolic:diracc}
[i\gamma^{\mu}\partial_{\mu}-m]\psi(x)=0 .
\end{equation}
The two quantities in (\ref{nikolic:a6}) 
\begin{equation}\label{nikolic:a6.1}
j^{\mu}_{\rm Dirac}=
\bar{\psi}(x)\gamma^{\mu}\psi(x) , 
\end{equation}
\begin{equation}\label{nikolic:a6.2}
 j^{\mu}=
\frac{i}{2}\psi^{\dagger}(x)  \!\stackrel{\leftrightarrow\;}{\partial^{\mu}}\!  \psi(x) ,
\end{equation}
are referred to as Dirac current and Klein-Gordon current, respectively.
They are both conserved
\begin{equation}\label{nikolic:a6.3}
 \partial_{\mu}j^{\mu}_{\rm Dirac}=0, \;\;\;\; 
\partial_{\mu}j^{\mu}=0.
\end{equation}
The first conservation is a consequence of (\ref{nikolic:diracc}), while
the second conservation is a consequence of (\ref{nikolic:KGsp}).

\subsubsection{Many particles with spin $\frac{1}{2}$}

The wave function for $n$ particles with spin $\frac{1}{2}$ has the form
$\psi_{l_1\dots l_n}(x_1,\ldots,x_n)$, where each $l_a$ is a spinor index.
It satisfies $n$ Dirac equations. A convenient way to write them is
\begin{equation}\label{nikolic:diraccn}
[i\gamma^{\mu}_a\partial_{a\mu}-m]\psi=0 ,
\end{equation}
where $\gamma^{\mu}_a$ is a ``matrix'' with $2n$ indices
\begin{equation}\label{nikolic:gamman}
 (\gamma^{\mu}_a)_{l_1\dots l_n l'_1\dots l'_n}=
\delta_{l_1 l'_1} \cdots (\gamma^{\mu})_{l_a l'_a} \cdots \delta_{l_n l'_n} .
\end{equation}
In the more abstract language of direct products, we can also write (\ref{nikolic:gamman}) as
\begin{equation}
 \gamma^{\mu}_a=1\otimes\cdots \otimes\gamma^{\mu}\otimes\cdots\otimes 1. 
\end{equation}
Similarly, the wave function satisfies also $n$ Klein-Gordon equations
\begin{equation}\label{nikolic:KGspn}
 [\partial_a^{\mu}\partial_{a\mu}+m^2]\psi =0 .
\end{equation}
Consequently, there are $n$ conserved Klein-Gordon currents 
\begin{equation}\label{nikolic:curnsn}
j^{\mu}_a=\frac{i}{2}\psi^{\dagger} \!\stackrel{\leftrightarrow\;}{\partial^{\mu}_a}\! \psi ,
\end{equation}
\begin{equation}\label{nikolic:conssn}
\partial^{\mu}_a j_{a\mu}=0,
\end{equation}
which imply a single conservation equation
\begin{equation}\label{nikolic:conssns}
\sum_a\partial^{\mu}_a j_{a\mu}=0.
\end{equation}
A similar generalization of the Dirac current also exists, but we shall not need it.

\subsubsection{Particles with spin 1}
\label{nikolic:SECspin1}

The case of spin 1 is much simpler than the case of spin $\frac{1}{2}$. 
Consequently, we shall only briefly outline how spin 1 particles are described.

A 1-particle wave function is $\psi_{\alpha}(x)$ and carries one vector index ${\alpha}$.
It satisfies 4 equations (see, e.g., \cite{ryder})
\begin{equation}\label{nikolic:KGF}
\partial^{\alpha}F_{\alpha\beta}+m^2\psi_{\beta}=0 ,
\end{equation} 
where
\begin{equation}\label{nikolic:F}
F_{\alpha\beta}=\partial_{\alpha}\psi_{\beta}-\partial_{\beta}\psi_{\alpha} .
\end{equation}
By applying the derivative $\partial^{\beta}$ on (\ref{nikolic:KGF}), one finds
\begin{equation}\label{nikolic:Lorgauge}
 \partial_{\beta}\psi^{\beta}=0 .
\end{equation}
Therefore, (\ref{nikolic:KGF}) implies 4 Klein-Gordon equations
\begin{equation}\label{nikolic:KGs1p}
 [\partial^{\mu}\partial_{\mu}+m^2]\psi_{\alpha}(x) =0 .
\end{equation} 
However, (\ref{nikolic:KGF}) implies that not all 4 components $\psi_{\alpha}$ 
are independent. For example, the time-component can be expressed in terms of other
components as $\psi_0=-\partial^{\alpha}F_{\alpha 0}/m^2$. Therefore, the most
general positive-frequency solution of $(\ref{nikolic:KGF})$ can be written in the form
\begin{equation}\label{nikolic:e3.9fvec}
\psi^{\alpha}(x)=\int d^3k \sum_{l=1}^{3} \epsilon^{\alpha}_l({\bf k}) a_l({\bf k})
e^{-i[\omega({\bf k})x^0-{\bf k}{\bf x}]} , 
\end{equation}
which can be thought of as a generalization of (\ref{nikolic:e3.9f}).
Here $a_l({\bf k})$ are arbitrary functions, while $\epsilon^{\alpha}_l({\bf k})$
are fixed polarization vectors \cite{ryder}. Thus, a wave function is completely
determined by 3 independent functions $a_l({\bf k})$, $l=1,2,3$. 
This implies that the system can also be described by a 3-component wave function
\begin{equation}\label{nikolic:e3.9fvec3}
\psi_l(x)=\int d^3k \, a_l({\bf k})
e^{-i[\omega({\bf k})x^0-{\bf k}{\bf x}]} , 
\end{equation}
where all 3 components are independent.
Since each component of (\ref{nikolic:e3.9fvec3}) also satisfies the Klein-Gordon
equation, the Klein-Gordon current
\begin{equation}\label{nikolic:curs1p}
j^{\mu}=\frac{i}{2} \sum_l 
\psi^*_l \!\stackrel{\leftrightarrow\;}{\partial^{\mu}}\! \psi_l 
\end{equation}
is conserved
\begin{equation}\label{nikolic:curs1pcons}
 \partial_{\mu}j^{\mu}=0 .
\end{equation}
 
In the case on $n$ particles the wave function 
$\psi_{l_1\ldots l_n}(x_1,\ldots,x_n)$ carries $n$ polarization labels.
It satisfies $n$ Klein-Gordon equations
\begin{equation}\label{nikolic:KGs1}
 [\partial_a^{\mu}\partial_{a\mu}+m^2]\psi_{l_1\ldots l_n}(x_1,\ldots,x_n) =0 ,
\end{equation} 
so (\ref{nikolic:curs1p}) and (\ref{nikolic:curs1pcons}) generalize to
\begin{equation}\label{nikolic:curs1}
j_a^{\mu}=\frac{i}{2} \sum_{l_1,\ldots , l_n} \psi^*_{l_1\ldots l_n} 
\!\stackrel{\leftrightarrow\;}{\partial_a^{\mu}}\! \psi_{l_1\ldots l_n}\, ,
\end{equation}
\begin{equation}
 \partial_{\mu}j_a^{\mu}=0 ,
\end{equation}
which implies
\begin{equation}
\sum_a \partial_{a\mu}j_a^{\mu}=0 .
\end{equation}

The case $m=0$ is special, because this case describes a photon, the wave function
of which contains also a gauge symmetry.
Namely, the (1-particle) wave function satisfies the free Maxwell equation
\begin{equation}\label{nikolic:maxwell}
 \partial^{\alpha}F_{\alpha\beta}=0 ,
\end{equation}
which is invariant with respect to gauge transformations
\begin{equation}
 \psi_{\alpha}(x) \rightarrow \psi'_{\alpha}(x) =\psi_{\alpha}(x) +\partial_{\alpha}\Lambda(x) ,
\end{equation}
where $\Lambda(x)$ is an arbitrary function. 
This gauge freedom can be partially removed by
imposing the Lorentz-gauge condition (\ref{nikolic:Lorgauge}). However, when the gauge
freedom is removed completely, then only 2 independent physical (transverse) 
polarizations remain. Consequently, the equations above involving $l$-labels 
modify such that $l$ takes only 2 values $l=1,2$. A gauge transformation
can be reduced to a change of the polarization vectors $\epsilon^{\alpha}_l({\bf k})$.
Thus, unlike $\psi_{\alpha}(x)$,
the wave function $\psi_l(x)$ is gauge invariant. 

Finally note that, in the massless case, the wave function $\psi_{\alpha}(x)$ 
is {\em not} the electromagnetic vector potential $A_{\alpha}(x)$. The latter is real
(not complex), so is represented by a superposition of positive and negative frequencies.
The former is a superposition of positive frequencies only, so it cannot be real at all $x$.

\subsection{Bohmian interpretation}

Now we are finally ready to deal with the Bohmian interpretation of relativistic QM.
Of course, the Bohmian interpretation could also be introduced without a lot of the background described in the preceding sections, but with this background the Bohmian interpretation
is very natural and almost trivial.

We start from the observation that the quantum equation (\ref{nikolic:HJ}) has the same form
as the classical equation (\ref{nikolic:book22}), provided that we make the replacement
\begin{equation}\label{nikolic:u-q}
 U(x)\rightarrow\frac{m}{2}+Q(x) .
\end{equation}
The first term on the right-hand side of (\ref{nikolic:u-q})
is the classical potential (\ref{nikolic:u=m/2}), while the second
term is the quantum potential.\footnote{Recall that we work in units $\hbar=1$. 
In units in which $\hbar\neq 1$, it is easy to show that (\ref{nikolic:Q=}) attains an additional factor
$\hbar^2$, showing that the quantum potential $Q$ vanishes in the classical limit.}
This suggests the Bohmian interpretation, according to which (\ref{nikolic:HJ}) is the quantum
Hamilton-Jacobi equation and the particle has the trajectory given by (\ref{nikolic:book24})
\begin{equation}\label{nikolic:book24q}
 \frac{dX^{\mu}(s)}{ds}=-\frac{\partial^{\mu}S(X(s))}{m} .
\end{equation}
From (\ref{nikolic:book24q}), (\ref{nikolic:HJ}), and the identity
\begin{equation}
\frac{d}{ds}=\frac{dX^{\mu}}{ds}\partial_{\mu} ,
\end{equation}
one finds a quantum variant of (\ref{nikolic:book13})
\begin{equation}\label{nikolic:book13q}
 m\frac{d^2X^{\mu}(s)}{ds^2}=\partial^{\mu}Q(X(s)) .
\end{equation}

But is such motion of quantum particles consistent with the probabilistic predictions
studied in Secs.~\ref{nikolic:RQ1} and \ref{nikolic:RQ2}? We first observe that (\ref{nikolic:book24q})
can be written as
\begin{equation}\label{nikolic:book24q1}
 \frac{dX^{\mu}}{ds}=\frac{j^{\mu}}{m\psi^*\psi} \; ,
\end{equation}
where $j^{\mu}$ is given by (\ref{nikolic:loc2}). 
It is convenient to eliminate the factor $1/m$ by rescaling the parameter $s$, so that
(\ref{nikolic:book24q1}) becomes
\begin{equation}\label{nikolic:book24q2}
 \frac{dX^{\mu}}{ds}=V^{\mu} ,
\end{equation}
where
\begin{equation}\label{nikolic:V}
V^{\mu}=\frac{j^{\mu}}{\psi^*\psi} .
\end{equation}
Second, we observe that (\ref{nikolic:conserv}) can be written as
\begin{equation}\label{nikolic:conserv2}
 \partial_{\mu}(|\psi|^2V^{\mu})=0 .
\end{equation}
Since $\psi(x)$ does not explicitly depend on $s$, we also have a trivial identity
$\partial|\psi|^2 /\partial s =0$. Therefore (\ref{nikolic:conserv2}) can be written as
 \begin{equation}\label{nikolic:conss1}
\frac{\partial|\psi|^2}{\partial s} + \partial_{\mu}(|\psi|^2V^{\mu})=0 .
\end{equation}
This implies that the trajectories satisfying (\ref{nikolic:book24q2}) are consistent with the probabilistic 
interpretation (\ref{nikolic:e6}). Namely, if a statistical ensemble of particles has the distribution
(\ref{nikolic:e6}) of spacetime particle positions for some ``initial'' $s$, then (\ref{nikolic:conss1})
guarantees that this statistical ensemble has the distribution
(\ref{nikolic:e6}) for {\em any} $s$. 

This shows that particles have the same distribution of spacetime positions as
predicted by the purely probabilistic interpretation of QM. 
But what about other measurable quantities? For example,
what about the space distribution of particles described in purely probabilistic QM by (\ref{nikolic:e8})?
Or what about the statistical distribution of particle velocities? In general, 
in the Bohmian interpretation all these other quantities may have a distribution totally
different from those predicted by purely probabilistic QM. In particular, the Bohmian velocities
of particles may exceed the velocity of light (which occurs when the right-hand side of 
(\ref{nikolic:u-q}) becomes negative\footnote{Chapter 9. studies
a possible cosmological relevance of such faster-than-light velocities.}), 
while purely probabilistic QM does not allow such velocities
because the eigenstates $e^{-ip_{\mu}x^{\mu}}$ of the velocity operator
$\hat{p}_{\mu}/m$ are not solutions of (\ref{nikolic:KG}) for $p^{\mu}p_{\mu}<0$.
Yet, when a quantity is {\em measured}, then the two theories have {\em the same} 
measurable predictions. 
Namely, since the Bohmian interpretation is compatible with (\ref{nikolic:e6}), the probability
that the measuring apparatus will be found in the state $E_b(y)$ in (\ref{nikolic:meas2})
is given by (\ref{nikolic:meas6}), which is the same as that in the purely probabilistic interpretation.

Now the generalization to $n$ particles without spin is straightforward. Essentially, 
all equations above are rewritten such that each quantity having the index $\mu$ receives
an additional index $a$. In particular, Eqs.~(\ref{nikolic:book24q}), (\ref{nikolic:book13q}), (\ref{nikolic:book24q2}), 
(\ref{nikolic:V}), (\ref{nikolic:conss1})
generalize to
\begin{equation}\label{nikolic:book24qn}
 \frac{dX_a^{\mu}(s)}{ds}=-\frac{\partial_a^{\mu}S(X_1(s),\ldots,X_n(s))}{m} ,
\end{equation}
\begin{equation}\label{nikolic:book13qn}
 m\frac{d^2X_a^{\mu}(s)}{ds^2}=\partial_a^{\mu}Q(X_1(s),\ldots,X_n(s)) ,
\end{equation}
\begin{equation}\label{nikolic:book24q2n}
 \frac{dX_a^{\mu}}{ds}=V_a^{\mu} ,
\end{equation}
\begin{equation}\label{nikolic:Vn}
V_a^{\mu}=\frac{j_a^{\mu}}{\psi^*\psi} \; ,
\end{equation}
 \begin{equation}\label{nikolic:conss1n}
\frac{\partial|\psi|^2}{\partial s} + 
\sum_{a=1}^{n}\partial_{a\mu}(|\psi|^2 V_a^{\mu})=0 ,
\end{equation}
respectively.
In general, particles have nonlocal influences on each other, in exactly the same way
as in classical relativistic mechanics studied in Sec.~\ref{nikolic:Dynamics}.

Now let us generalize these results to particles with spin. When spin is present,
the analogy with the classical Hamilton-Jacobi equation is less useful. The crucial
requirement is the consistency with the purely probabilistic interpretation (\ref{nikolic:e6ng2}).
This is achieved by generalizing (\ref{nikolic:book24q2n}) and (\ref{nikolic:Vn}) to
\begin{equation}\label{nikolic:book24q2ns}
 \frac{dX_a^{\mu}}{ds}=V_a^{\mu} ,
\end{equation}
\begin{equation}\label{nikolic:Vns}
V_a^{\mu}=\frac{j_a^{\mu}}{\psi^{\dagger}\psi} \; ,
\end{equation}
where $j_a^{\mu}$ is a conserved current given by (\ref{nikolic:curnsn}) for
spin $\frac{1}{2}$ particles and (\ref{nikolic:curs1}) for spin 1 particles. 
The compatibility with (\ref{nikolic:e6ng2}) is provided by the generalization of (\ref{nikolic:conss1n})
\begin{equation}\label{nikolic:conss1ns}
\frac{\partial\psi^{\dagger}\psi}{\partial s} + 
\sum_{a=1}^{n}\partial_{a\mu}(\psi^{\dagger}\psi V_a^{\mu})=0 .
\end{equation}

\section{Quantum field theory} 

\subsection{Main ideas of QFT and its Bohmian interpretation}
\label{nikolic:SEC2}

So far, we have been considering systems with a fixed number $n$ of particles.
However, in many physical systems the number of particles is not fixed.
Instead, particles may be created or destroyed. To describe such processes,
a more general formalism is needed. This formalism is known as
{\it quantum field theory} (QFT).

The simplest way to understand the kinematics of QFT is as follows.
Let ${\cal H}^{(n)}$ denote the Hilbert space associated with quantum mechanics
of a fixed number $n$ of particles, where $n\geq 1$. 
An element of this Hilbert space is a quantum state of $n$ particles, denoted
abstractly by $|n\rangle$.
In fact, the case $n=0$ can also be included, by defining a new
trivial 1-dimensional Hilbert space ${\cal H}^{(0)}$. This trivial space has only
1 linearly independent element denoted by $|0\rangle$, which represents
the vacuum, i.e., the state with no particles. From all these Hilbert spaces
one can construct a single Hilbert space ${\cal H}$ containing all of them
as subspaces, through a direct sum
\begin{equation}
 {\cal H}=\bigoplus_{n=0}^{\infty}{\cal H}^{(n)}
\equiv {\cal H}^{(0)} \oplus {\cal H}^{(1)} \oplus {\cal H}^{(2)} \oplus \cdots .
\end{equation}
QFT is nothing but the theory of states in the Hilbert space ${\cal H}$.
A general state in this space is a linear combination of the form
 \begin{equation}\label{nikolic:QFTPsi}
  |\Psi\rangle = \sum_{n=0}^{\infty} c_n |n\rangle .
 \end{equation}
QFT is the theory of states $(\ref{nikolic:QFTPsi})$.\footnote{In such a view of QFT, 
the fundamental physical objects are particles, while fields only play an auxiliary role.
There is also a different view of QFT in which fields play a more fundamental role
than particles.
An example of such a different view is presented in Chapter 9. However, in the context of 
Bohmian interpretation, there are at least two problems when fields are viewed 
as being more fundamental.
First, it is not known how to make the Bohmian equations of motion for 
bosonic fields relativistic covariant.
Second, it is not known how to include the fermionic fields. Various proposals
for solving these two problems exist, but none of them seems completely satisfying.
On the other hand, 
we shall see that such problems can be solved in a 
simple and natural way when the Bohmian interpretation
is based on particles.} 

As a simple example, consider a QFT state of the form 
\begin{equation}\label{nikolic:e2.1}
|\Psi\rangle = |1\rangle + |2\rangle ,
\end{equation}
which is a superposition of a 1-particle state $|1\rangle$ and a 2-particle state
$|2\rangle$. For example, it may represent an unstable particle for which 
we do not know if it has already decayed into 2 new particles (in which case it is
described by $|2\rangle$) or has not decayed yet (in which case it is
described by $|1\rangle$). However, it is known that
one always observes either one unstable particle
(the state $|1\rangle$) or two decay products (the state $|2\rangle$). One never 
observes the superposition (\ref{nikolic:e2.1}). Why?

To answer this question, let us try with a Bohmian approach. 
One can associate a 1-particle wave function
$\Psi_1(x_1)$ with the state $|1\rangle$ and a 2-particle  wave function
$\Psi_2(x_2,x_3)$ with the state $|2\rangle$, where $x_A$ is the spacetime
position of the particle labeled by $A=1,2,3$.
Then the state (\ref{nikolic:e2.1}) is represented by a superposition
\begin{equation}\label{nikolic:e2.2}
\Psi(x_1,x_2,x_3) = \Psi_1(x_1) + \Psi_2(x_2,x_3) .
\end{equation} 
However, the Bohmian interpretation of such a superposition will describe
{\em three} particle trajectories. On the other hand, we should observe
either one or two particles, not three particles. How to explain that?

The key is to take into account the properties of the {\em measuring apparatus}. 
If the number of particles is measured, then instead of (\ref{nikolic:e2.2}) we actually have
a wave function of the form
\begin{equation}\label{nikolic:e2.5}
\Psi(x_1,x_2,x_3,y) = \Psi_1(x_1)E_1(y) + \Psi_2(x_2,x_3)E_2(y) .
\end{equation}
The detector wave functions $E_1(y)$ and $E_2(y)$ do not overlap. 
Hence, if $y$ takes a value $Y$ in the support of $E_2$, then this value is not
in the support of $E_1$, i.e., $E_1(Y)=0$. Consequently, the motion of the
measured particles is described by the conditional wave function 
$\Psi_2(x_2,x_3)E_2(Y)$.
The effect is the same as if (\ref{nikolic:e2.2}) collapsed to $\Psi_2(x^2,x^3)$.

Now, what happens with the particle having the spacetime position $x_1$?
In general, its motion in spacetime may be expected to be
described by the relativistic Bohmian equation of motion 
\begin{equation}\label{nikolic:e2.6}
\frac{dX^{\mu}_1(s)}{ds} = \frac{ 
\frac{i}{2} \Psi^*\!\stackrel{\leftrightarrow\;}{\partial^{\mu}_1}\! \Psi }{\Psi^*\Psi} .
\end{equation}
However,
if the absence of the overlap between $E_1(y)$ and $E_2(y)$ is exact, then
the effective wave function does not depend on $x_1$, i.e., the derivatives in
(\ref{nikolic:e2.6}) vanish. Consequently, all 4 components of the 4-velocity (\ref{nikolic:e2.6})
are zero. The particle does not change its spacetime position $X^{\mu}_1$. 
It is an object without an extension not only in space, but also in time. 
It can be thought of as a pointlike particle that exists only at one instant of time
$X^{0}_1$. It lives too short to be detected. Effectively, this particle 
behaves as if it did not exist at all.

Now consider a more realistic variation of the measuring procedure, taking into account
the fact that the measured particles become entangled with the measuring apparatus 
at some finite time $T$.
Before that, the wave function of the measured particles is really well described by 
(\ref{nikolic:e2.2}). Thus, before the interaction with the measuring apparatus, all 3 
particles described by (\ref{nikolic:e2.2}) have continuous trajectories in spacetime.
All 3 particles exist. But at time $T$, the total wave function significantly changes.
Either (i) $y$ takes a value from the support of $E_2$ in which case
$dX_1^{\mu}/ds$ becomes zero, or (ii) $y$ takes a value from the support of $E_1$
in which case  $dX_2^{\mu}/ds$ and $dX_3^{\mu}/ds$ become zero.
After time $T$, either the particle 1 does not longer change its spacetime position,
or the particles 2 and 3 do not longer change their spacetime positions.
The trajectory of the particle 1 or the trajectories of the particles 2 and 3 terminate
at $T$, i.e., they do not exist for times $t>T$. This is how relativistic Bohmian interpretation
describes the particle destruction. 

Unfortunately, the mechanism above works only in a very special case
in which the absence of the overlap between $E_1(y)$ and $E_2(y)$ is exact.
In a more realistic situation this overlap is negligibly small, but not exactly zero.
In such a situation neither of the particles will have exactly zero 4-velocity.
Consequently, neither of the particles will be really destroyed.
Nevertheless, the measuring apparatus will still behave as if some particles 
have been destroyed. For example, if $y$ takes value $Y$ 
for which $E_1(Y)\ll E_2(Y)$, then for all practical purposes the measuring apparatus
behaves as if the wave function collapsed to the second term in (\ref{nikolic:e2.5}).
The particles with positions $X_2$ and $X_3$ also behave in that way. 
Therefore, even though the particle with the position $X_1$ is not really destroyed,
an effective wave-function collapse still takes place. The influence of the particle
with the position $X_1$ on the measuring apparatus described by $Y$
is negligible, which is effectively the same 
as if this particle has been destroyed.

Of course, the interaction with the measuring apparatus is not the only mechanism
that may induce destruction of particles. Any interaction with the environment
may do that.  Or more generally, any interactions among particles 
may induce not only particle destruction, but also particle creation. 
Whenever the wave function $\Psi(x_1,x_2,x_3,x_4, \ldots )$ does not really vary
(or when this variation is negligible)
with some of $x_A$ for some range of values of $x_A$,
then at the edge of this range a trajectory of the particle $A$ may exhibit
true (or apparent) creation or destruction. 

In general, a QFT state may be a superposition of $n$-particle states
with $n$ ranging from $0$ to $\infty$. Thus, $\Psi(x_1,x_2,x_3,x_4, \ldots )$
should be viewed as a function that lives in the space of infinitely many coordinates
$x_A$, $A=1,2,3,4,\ldots, \infty$. In particular, the 1-particle wave function
$\Psi_1(x_1)$ should be viewed as a function $\Psi_1(x_1,x_2, \ldots)$
with the property $\partial^{\mu}_A \Psi_1 =0$ for $A=2,3,\ldots, \infty$.
It means that any wave function in QFT describes an infinite number of particles,
even if most of them have zero 4-velocity. As we have already explained, particles
with zero 4-velocity are dots in spacetime. The initial spacetime position of any particle
may take any value, with the probability proportional to
$|\Psi_1(x_1,x_2, \ldots)|^2$. In addition to one continuous particle trajectory, there is also an infinite
number of ``vacuum'' particles which live for an infinitesimally short time.

The purpose of the remaining subsections of this section is to further elaborate
the ideas presented in this subsection 
and to put them into a more precise framework.

\subsection{Measurement in QFT as entanglement with the environment}
\label{nikolic:SEC3.1}

Let $\{|b\rangle \}$ be some orthonormal basis of 1-particle states.
A general normalized 1-particle state is
\begin{equation}\label{nikolic:e3.1}
 |\Psi_1\rangle = \sum_b c_b |b\rangle ,
\end{equation}
where the normalization condition implies $ \sum_b |c_b|^2=1$. 
From the basis $\{|b\rangle \}$ one can construct the $n$-particle basis
$\{|b_1,\ldots, b_n\rangle \}$, where
\begin{equation}\label{nikolic:e3.2}
|b_1,\ldots, b_n\rangle = S_{\{b_1,\ldots, b_n\}}|b_1\rangle \cdots  |b_n\rangle .
\end{equation}
Here $S_{\{b_1,\ldots, b_n\}}$ denotes the symmetrization over all $\{b_1,\ldots, b_n\}$
for bosons, or antisymmetrization for fermions. The most general state in QFT
describing these particles can be written as
\begin{equation}\label{nikolic:e3.3}
 |\Psi\rangle = c_0 |0\rangle +
\sum_{n=1}^{\infty} \sum_{b_1,\ldots, b_n} c_{n;b_1,\ldots,b_n} 
|b_1,\ldots, b_n\rangle ,
\end{equation}
where the vacuum $|0\rangle$ is also introduced. 
Now the normalization condition implies
$|c_0|^2+\sum_{n=1}^{\infty} \sum_{b_1,\ldots, b_n} |c_{n;b_1,\ldots,b_n}|^2 =1$.

Now let as assume that the number of particles is measured. It implies that the particles
become entangled with the environment, such that the total state describing both 
the measured particles and the environment takes the form
\begin{equation}\label{nikolic:e3.4}
 |\Psi\rangle_{\rm total}  =  c_0 |0\rangle |E_0\rangle 
 + \sum_{n=1}^{\infty} \sum_{b_1,\ldots, b_n} c_{n;b_1,\ldots,b_n}  
|b_1,\ldots, b_n\rangle |E_{n;b_1,\ldots,b_n} \rangle .
\end{equation}
The environment states $|E_0\rangle$, $|E_{n;b_1,\ldots,b_n} \rangle$ are
macroscopically distinct. They describe what the observers really observe.
When an observer observes that the environment is in the state
$|E_0\rangle$ or $|E_{n;b_1,\ldots,b_n} \rangle$, then one says that the 
original measured QFT state is in the state $|0\rangle$ or 
$|b_1,\ldots, b_n\rangle$, respectively. In particular, this is how the number of 
particles is measured in a state (\ref{nikolic:e3.3}) with an uncertain number of particles.
The probability that the environment will be found in the state
$|E_0\rangle$ or $|E_{n;b_1,\ldots,b_n} \rangle$ is 
equal to $|c_0|^2$ or $|c_{n;b_1,\ldots,b_n}|^2$, respectively.

Of course, (\ref{nikolic:e3.3}) is not the only way the state $|\Psi\rangle$ can be expanded.
In general, it can be expanded as
\begin{equation}\label{nikolic:e3.5}
 |\Psi\rangle = \sum_{\xi} c_{\xi} |\xi\rangle ,
\end{equation}
where $|\xi\rangle$ are some normalized (not necessarily orthogonal)
states that do not need to have a definite number of particles.
A particularly important example are coherent states (see, e.g., \cite{bal}),
which minimize the products of uncertainties of fields and their 
canonical momenta. Each coherent state is a superposition
of states with all possible numbers of particles, including zero. 
The coherent states are overcomplete and not orthogonal. Yet, the
expansion (\ref{nikolic:e3.5}) may be an expansion in terms of coherent states $|\xi\rangle$
as well. 

Furthermore, the entanglement with the environment does not
necessarily need to take the form (\ref{nikolic:e3.4}). Instead, it may take a
more general form
\begin{equation}\label{nikolic:e3.6}
 |\Psi\rangle_{\rm total}  = \sum_{\xi} c_{\xi} |\xi\rangle |E_\xi\rangle , 
\end{equation}
where $|E_\xi\rangle$ are macroscopically distinct. 
In principle, the interaction with the environment may create the entanglement
(\ref{nikolic:e3.6}) with respect to any set of states $\{ |\xi\rangle \}$.
In practice, however, some types of expansions are preferred.
This fact can be explained by the theory of decoherence \cite{schloss}, which 
explains why states of the form of (\ref{nikolic:e3.6}) are stable only for some particular
sets $\{ |\xi\rangle \}$. In fact, depending on details of the interactions
with the environment, in most real situations the entanglement takes either the form
(\ref{nikolic:e3.4}) or the form (\ref{nikolic:e3.6}) with coherent states $|\xi\rangle$.
Since coherent states minimize the uncertainties of fields and their canonical momenta,
they behave very much like classical fields. This explains why experiments in quantum
optics can often be better described in terms of fields rather than particles 
(see, e.g., \cite{bal}). In fact, the theory of decoherence can explain 
under what conditions the coherent-state basis becomes preferred over
basis with definite numbers of particles \cite{zeh,zurek}. 

Thus, decoherence induced by interaction with the environment
can explain why do we observe either a definite number
of particles or coherent states that behave very much like classical fields.
However, decoherence alone 
cannot explain why do we observe 
some particular state of definite number of particles and not some other, 
or why do we observe some particular coherent state and not some other.
Instead, a possible explanation is provided by the Bohmian interpretation.

\subsection{Free scalar QFT in the particle-position picture}
\label{nikolic:SEC3.2}

The purpose of this subsection is to see in detail
how states of free QFT without spin can be represented
by wave functions. They include wave functions with definite number of
particles (discussed in Sec.~\ref{nikolic:RQ}), as well as their superpositions.

Consider a free scalar hermitian field operator $\hat{\phi}(x)$ satisfying the 
Klein-Gordon equation
\begin{equation}\label{nikolic:e3.7}
 \partial^{\mu}\partial_{\mu}\hat{\phi}(x)+ m^2\hat{\phi}(x) =0. 
\end{equation}
The field can be decomposed as
\begin{equation}\label{nikolic:e3.8}
 \hat{\phi}(x)=\hat{\psi}(x)+\hat{\psi}^{\dagger}(x) ,
\end{equation}
where $\hat{\psi}$ and $\hat{\psi}^{\dagger}$ can be expanded as
\begin{eqnarray}\label{nikolic:e3.9}
& \hat{\psi}(x)=\displaystyle\int d^3k \, f({\bf k}) \, \hat{a}({\bf k})
e^{-i[\omega({\bf k})x^0-{\bf k}{\bf x}]} , &
\nonumber \\
& \hat{\psi}^{\dagger}(x)=\displaystyle\int d^3k \, f({\bf k}) \, \hat{a}^{\dagger}({\bf k})
e^{i[\omega({\bf k})x^0-{\bf k}{\bf x}]} . &
\end{eqnarray}
Here 
\begin{equation}\label{nikolic:e3.9'}
\omega({\bf k})=\sqrt{{\bf k}^2+m^2}
\end{equation} 
is the $k_0$ component
of the 4-vector $k=\{ k_{\mu} \}$, and
$\hat{a}^{\dagger}({\bf k})$ and $\hat{a}({\bf k})$
are the creation and destruction operators, respectively (see, e.g., \cite{bd2}),
satisfying the commutation relations
$[\hat{a}({\bf k}),\hat{a}({\bf k}')]=
[\hat{a}^{\dagger}({\bf k}),\hat{a}^{\dagger}({\bf k}')]=0$,
$[\hat{a}({\bf k}),\hat{a}^{\dagger}({\bf k}')] \propto \delta^3({\bf k}-{\bf k}')$.
The function
$f({\bf k})$ is a real positive function which we do not specify explicitly
because several different choices appear in the literature, corresponding to
several different choices of normalization. All subsequent equations will
be written in forms that do not explicitly depend on this choice.

We define the operator
\begin{equation}\label{nikolic:e3.10}
 \hat{\psi}_n(x_{n,1}, \ldots , x_{n,n}) = d_n 
S_{ \{x_{n,1}, \ldots , x_{n,n} \} } 
\hat{\psi}(x_{n,1}) \cdots \hat{\psi}(x_{n,n}) .
\end{equation}
The symbol $ S_{ \{x_{n,1}, \ldots , x_{n,n} \} }$ denotes the symmetrization, 
reminding us that the expression
is symmetric under the exchange of coordinates $\{x_{n,1}, \ldots , x_{n,n} \}$.
(Note, however, that the product of operators on the right hand side of (\ref{nikolic:e3.10})
is in fact automatically symmetric because 
the operators $\hat{\psi}(x)$ commute, i.e., $[\hat{\psi}(x),\hat{\psi}(x')]=0$.) 
The parameter $d_n$ is a normalization constant determined by the normalization condition
that will be specified below.
The operator (\ref{nikolic:e3.10}) allows us to define $n$-particle states
in the basis of particle spacetime positions, as
\begin{equation}\label{nikolic:e3.11}
 |x_{n,1}, \ldots , x_{n,n}\rangle = \hat{\psi}^{\dagger}_n(x_{n,1}, \ldots , x_{n,n})
|0\rangle .
\end{equation}

The normalization function $f({\bf k})$ in (\ref{nikolic:e3.9}) can be chosen such that
all states of the form (\ref{nikolic:e3.11}) at a fixed common time 
$x^0_{n,1}= \cdots =x^0_{n,n}=t$, 
together with the vacuum $|0\rangle$, form
a complete and orthogonal basis in the Hilbert space of physical states.
For example, for 1-particle states the orthogonality relation reads
$\langle {\bf x};t|{\bf x}';t\rangle=\delta^3({\bf x}-{\bf x}')$,
and similarly for $n$-particle states. However, for such a choice of 
$f({\bf k})$, the operators (\ref{nikolic:e3.9}) are not Lorentz invariant.
Thus, it is more appropriate to sacrifice orthogonality by
choosing $f({\bf k})$ such that (\ref{nikolic:e3.9}) are Lorentz invariant. 
In the rest of the analysis we assume such a Lorentz-invariant
normalization of (\ref{nikolic:e3.9}). 

If $|\Psi_n\rangle$ is an arbitrary (but normalized) $n$-particle state,
then this state can be represented by the $n$-particle wave function
\begin{equation}\label{nikolic:e3.12}
 \psi_n(x_{n,1}, \ldots , x_{n,n}) = \langle x_{n,1}, \ldots , x_{n,n} |\Psi_n\rangle .
\end{equation}
We also have
\begin{equation}\label{nikolic:e3.12.1}
 \langle x_{n,1}, \ldots , x_{n,n} |\Psi_{n'}\rangle =0 \;\;{\rm for}\;\; n\neq n' .
\end{equation}
We choose the normalization constant $d_n$ in (\ref{nikolic:e3.10}) such that the following
normalization condition is satisfied
\begin{equation}\label{nikolic:e3.13}
 \int d^4x_{n,1}\cdots \int d^4x_{n,n} \, | \psi_n(x_{n,1}, \ldots , x_{n,n})|^2 =1 .
\end{equation}
However, this implies that the wave functions 
$\psi_n(x_{n,1}, \ldots , x_{n,n})$ and $\psi_{n'}(x_{n',1}, \ldots , x_{n',n'})$,
with different values of $n$ and $n'$,
are normalized in different spaces. On the other hand, we want these wave functions
to live in the same space, such that we can form superpositions of wave functions
describing different numbers of particles. To accomplish this, we define
\begin{equation}\label{nikolic:e3.14}
\Psi_n(x_{n,1}, \ldots , x_{n,n})=\sqrt{ \frac{{\cal V}^{(n)}}{{\cal V}} } \,
\psi_n(x_{n,1}, \ldots , x_{n,n}) ,
\end{equation}
where
\begin{equation}\label{nikolic:e3.15}
 {\cal V}^{(n)}=\int d^4x_{n,1}\cdots \int d^4x_{n,n} ,
\end{equation}
\begin{equation}\label{nikolic:e3.16}
 {\cal V}=\prod_{n=1}^{\infty} {\cal V}^{(n)} ,
\end{equation}
are volumes of the corresponding configuration spaces.
In particular, the wave function of the vacuum is
\begin{equation}\label{nikolic:e3.17}
\Psi_0=\frac{1}{\sqrt{{\cal V}}} .
\end{equation}
This provides that all wave functions are normalized in the same configuration space
as
 \begin{equation}\label{nikolic:e3.18}
\int {\cal D}\vec{x} \, | \Psi_n(x_{n,1}, \ldots , x_{n,n})|^2 =1 ,
\end{equation}
where we use the notation
\begin{equation}\label{nikolic:e3.19}
 \vec{x}=(x_{1,1},x_{2,1},x_{2,2},\ldots ),
\end{equation}
\begin{equation}\label{nikolic:e3.20}
 {\cal D}\vec{x} = \prod_{n=1}^{\infty} \, \prod_{a_{n}=1}^{n} d^4x_{n,a_{n}} .
\end{equation}

Note that the physical Hilbert space does not contain non-symmetrized
states, such as a 3-particle state $|x_{1,1}\rangle |x_{2,1},x_{2,2}\rangle$.
It also does not contain states that do not satisfy (\ref{nikolic:e3.9'}). 
Nevertheless, the notation can be further simplified by introducing an extended
kinematic Hilbert space that contains such unphysical states as well.
Every physical state can be viewed as a state in such an extended Hilbert space,
although most of the states in the extended Hilbert space are not physical.
In this extended space it is convenient to denote the pair of
labels $(n,a_n)$ by a single label $A$. Hence,  (\ref{nikolic:e3.19}) and (\ref{nikolic:e3.20})
are now written as
\begin{equation}\label{nikolic:e3.21}
 \vec{x}=(x_1,x_2,x_3,\ldots ),
\end{equation}
\begin{equation}\label{nikolic:e3.22}
 {\cal D}\vec{x} = \prod_{A=1}^{\infty} d^4x_A .
\end{equation}
Similarly, (\ref{nikolic:e3.16}) with (\ref{nikolic:e3.15}) is now written as
\begin{equation}\label{nikolic:e3.23}
 {\cal V}=\int \prod_{A=1}^{\infty} d^4x_A .
\end{equation}
The particle-position basis of this extended space is denoted by $|\vec{x})$ (which should be
distinguished from $|\vec{x}\rangle$ which would denote a symmetrized state
of an infinite number of physical particles).  
Such a basis allows us to write  
the physical wave function (\ref{nikolic:e3.14}) as a wave function 
on the extended space 
\begin{equation}\label{nikolic:e3.25}
 \Psi_n(\vec{x})=(\vec{x}|\Psi_n\rangle .
\end{equation}
Now (\ref{nikolic:e3.18}) takes a simpler form 
\begin{equation}\label{nikolic:e3.18'}
\int {\cal D}\vec{x} \, | \Psi_n(\vec{x})|^2 =1 .
\end{equation}
The unit operator on the extended space is
\begin{equation}\label{nikolic:e3.24}
 1=\int {\cal D}\vec{x} \, |\vec{x}) (\vec{x}| ,
\end{equation}
while the scalar product is
\begin{equation}\label{nikolic:e3.24'}
 (\vec{x}|\vec{x}')=\delta(\vec{x}-\vec{x}') ,
\end{equation}
with $\delta(\vec{x}-\vec{x}') \equiv \prod_{A=1}^{\infty}\delta^4(x_A-x'_A)$.
A general physical state can be written as
\begin{equation}\label{nikolic:e3.26}
 \Psi(\vec{x})=(\vec{x}|\Psi\rangle = 
\sum_{n=0}^{\infty}c_n \Psi_n(\vec{x}) .
\end{equation}
It is also convenient to write this as
\begin{equation}\label{nikolic:e3.27}
\Psi(\vec{x})=\sum_{n=0}^{\infty} \tilde{\Psi}_n(\vec{x}) ,
\end{equation}
where the tilde denotes a wave function that is not necessarily normalized.
The total wave function is normalized, in the sense that
\begin{equation}\label{nikolic:e3.26.1}
 \int {\cal D}\vec{x} \, |\Psi(\vec{x})|^2=1 ,
\end{equation}
implying
\begin{equation}\label{nikolic:e3.26.2}
 \sum_{n=0}^{\infty}|c_n|^2=1 .
\end{equation}

Next, we introduce the operator
\begin{equation}\label{nikolic:e3.28}
 \Box = \sum_{A=1}^{\infty} \partial_A^{\mu} \partial_{A\mu} .
\end{equation}
From the equations above (see, in particular, (\ref{nikolic:e3.7})-(\ref{nikolic:e3.12})), 
it is easy to show that $\Psi_n(\vec{x})$ satisfies
\begin{equation}\label{nikolic:e3.29}
 \Box\Psi_n(\vec{x}) +nm^2 \Psi_n(\vec{x}) =0 .
\end{equation}
Introducing a hermitian number-operator $\hat{N}$ with the property
\begin{equation}\label{nikolic:e3.30}
 \hat{N}\Psi_n(\vec{x}) = n \Psi_n(\vec{x}) ,
\end{equation}
one finds that a general physical state (\ref{nikolic:e3.26}) satisfies the generalized
Klein-Gordon equation
\begin{equation}\label{nikolic:e3.31}
 \Box\Psi(\vec{x})+m^2\hat{N}\Psi(\vec{x})=0.
\end{equation}
We also introduce the generalized Klein-Gordon current
\begin{equation}\label{nikolic:e3.32}
 J^{\mu}_A(\vec{x})=
\frac{i}{2} \Psi^*(\vec{x})\!\stackrel{\leftrightarrow\;}{\partial^{\mu}_A}\! \Psi (\vec{x}) .
\end{equation}
From (\ref{nikolic:e3.31}) one finds that, in general, this current is not conserved
\begin{equation}\label{nikolic:e3.33}
 \sum_{A=1}^{\infty}\partial_{A\mu}J^{\mu}_A(\vec{x}) = J(\vec{x}) ,
\end{equation}
where
\begin{equation}\label{nikolic:e3.34}
 J(\vec{x})=-\frac{i}{2}m^2 
\Psi^*(\vec{x})\!\stackrel{\;\leftrightarrow}{\hat{N}}\! \Psi (\vec{x}) ,
\end{equation}
and $\Psi' \!\stackrel{\;\leftrightarrow}{\hat{N}}\! \Psi \equiv 
\Psi' (\hat{N}\Psi) - ( \hat{N} \Psi') \Psi$. 
From (\ref{nikolic:e3.34}) we see that the current is conserved in two special cases:
(i) when $\Psi=\Psi_n$ (a state with a definite number of physical
particles), or (ii) when $m^2=0$ (any physical state of massless particles). 

Finally, let us rewrite some of the
main results of this (somewhat lengthy) subsection in a form
that will be suitable for a generalization in the next subsection.
A general physical state can be written in the form
\begin{equation}\label{nikolic:s1}
|\Psi\rangle = \sum_{n=0}^{\infty} c_n |\Psi_n\rangle = 
\sum_{n=0}^{\infty}  |\tilde{\Psi}_n\rangle .
\end{equation} 
The corresponding unnormalized $n$-particle wave functions are
\begin{equation}\label{nikolic:s2}
 \tilde{\psi}_n(x_{n,1},\ldots,x_{n,n}) =
\langle 0|\hat{\psi}_n(x_{n,1},\ldots,x_{n,n})|\Psi\rangle .
\end{equation}
There is a well-defined transformation 
\begin{equation}\label{nikolic:s3}
 \tilde{\psi}_n(x_{n,1},\ldots,x_{n,n}) \rightarrow \tilde{\Psi}_n(\vec{x})  
\end{equation}
from the physical Hilbert space to the extended Hilbert space, so that
the general state (\ref{nikolic:s1}) can be represented by a single wave function
\begin{equation}\label{nikolic:s4}
\Psi(\vec{x})=\sum_{n=0}^{\infty} c_n \Psi_n(\vec{x})
=\sum_{n=0}^{\infty} \tilde{\Psi}_n(\vec{x}).
\end{equation}

\subsection{Generalization to interacting QFT}
\label{nikolic:SEC3.3}

In this subsection we discuss the generalization of the results of the preceding subsection
to the case in which the field operator $\hat{\phi}$ does not satisfy the free 
Klein-Gordon equation (\ref{nikolic:e3.7}). For example,
if the classical action for the field is
\begin{equation}\label{nikolic:e3.44}
 S=\int d^4x  \left[ \frac{1}{2}(\partial^{\mu}\phi) (\partial_{\mu}\phi)
-\frac{m^2}{2}\phi^2 - \frac{\lambda}{4}\phi^4 \right] ,
\end{equation}
then (\ref{nikolic:e3.7}) generalizes to
 \begin{equation}\label{nikolic:e3.45}
 \partial^{\mu}\partial_{\mu}\hat{\phi}_H(x)+ 
m^2\hat{\phi}_H(x) +\lambda \hat{\phi}_H^3(x)=0, 
\end{equation}
where $\hat{\phi}_H(x)$ is the field operator in the Heisenberg picture.
(From this point of view, the operator $\hat{\phi}(x)$ 
defined by (\ref{nikolic:e3.8}) and (\ref{nikolic:e3.9}) and satisfying the free Klein-Gordon 
equation (\ref{nikolic:e3.7}) is the field operator in the interaction (Dirac) picture.)
Thus, instead of (\ref{nikolic:s2}) now we have
\begin{equation}\label{nikolic:s2'}
 \tilde{\psi}_n(x_{n,1},\ldots,x_{n,n}) =
\langle 0|\hat{\psi}_{nH}(x_{n,1},\ldots,x_{n,n})|\Psi\rangle ,
\end{equation}
where $|\Psi\rangle$ and $|0\rangle$ are states in the Heisenberg picture.
Assuming that (\ref{nikolic:s2'}) has been calculated (we shall see below how
in practice it can be done), the rest of the job is straightforward.
One needs to make the transformation (\ref{nikolic:s3}) in the same way
as in the free case, which leads to an interacting variant of  (\ref{nikolic:s4})
\begin{equation}\label{nikolic:s4'}
\Psi(\vec{x})=\sum_{n=0}^{\infty} \tilde{\Psi}_n(\vec{x}) .
\end{equation}
The wave function (\ref{nikolic:s4'}) encodes the complete information about the
properties of the interacting system. 

Now let us see how (\ref{nikolic:s2'}) can be calculated in practice. Any operator
$\hat{O}_H(t)$ in the Heisenberg picture depending on a single time-variable $t$
can be written in terms of operators in the interaction picture as 
\begin{equation}\label{nikolic:e3.46}
\hat{O}_H(t)=\hat{U}^{\dagger}(t)\hat{O}(t)\hat{U}(t) , 
\end{equation}
where
\begin{equation}\label{nikolic:e3.47}
 \hat{U}(t)=Te^{-i\int_{t_0}^t dt' \hat{H}_{\rm int}(t')} ,
\end{equation}
$t_0$ is some appropriately chosen ``initial'' time, $T$ denotes the time ordering,
and $\hat{H}_{\rm int}$
is the interaction part of the Hamiltonian expressed as a functional of field operators
in the interaction picture (see, e.g., \cite{chengli}). 
For example, for the action (\ref{nikolic:e3.44}) we have
\begin{equation}\label{nikolic:e3.48}
 \hat{H}_{\rm int}(t)=\frac{\lambda}{4} \int d^3x \, :\!\hat{\phi}^4({\bf x},t)\!: ,
\end{equation}
where $:\;:$ denotes the normal ordering.
The relation (\ref{nikolic:e3.46}) can be inverted, leading to
\begin{equation}\label{nikolic:e3.49}
\hat{O}(t)=\hat{U}(t)\hat{O}_H(t)\hat{U}^{\dagger}(t) . 
\end{equation}
Thus, the relation (\ref{nikolic:e3.10}), which is now valid in the interaction picture,
allows us to write an analogous relation in the Heisenberg picture 
\begin{equation}\label{nikolic:e3.10'}
 \hat{\psi}_{nH}(x_{n,1}, \ldots , x_{n,n})  =  d_n 
S_{ \{x_{n,1}, \ldots , x_{n,n} \} } 
 \hat{\psi}_H(x_{n,1}) \cdots \hat{\psi}_H(x_{n,n}) ,
\end{equation}
where
\begin{equation}\label{nikolic:e3.50}
\hat{\psi}_H(x_{n,a_n})=\hat{U}^{\dagger}(x^0_{n,a_n})
\hat{\psi}(x_{n,a_n})\hat{U}(x^0_{n,a_n}) .
\end{equation}
By expanding (\ref{nikolic:e3.47}) in powers of $\int_{t_0}^t dt' \hat{H}_{\rm int}$,
this allows us to calculate (\ref{nikolic:e3.10'}) and (\ref{nikolic:s2'}) perturbatively.
In (\ref{nikolic:s2'}), the states in the Heisenberg picture $|\Psi\rangle$ and $|0\rangle$
are identified with the states in the interaction picture at the initial time
$|\Psi(t_0)\rangle$ and $|0(t_0)\rangle$, respectively.

To demonstrate that such a procedure leads to a physically sensible result,
let us see how it works in the special (and more familiar) case of the equal-time
wave function. It is given by $\tilde{\psi}_n(x_{n,1},\ldots,x_{n,n})$ 
calculated at $x^0_{n,1}=\cdots=x^0_{n,n}\equiv t$.
Thus, (\ref{nikolic:s2'}) reduces to
\begin{equation}\label{nikolic:e3.51}
 \tilde{\psi}_n({\bf x}_{n,1},\ldots,{\bf x}_{n,n};t) = d_n
\langle 0(t_0)|  \hat{U}^\dagger(t) \hat{\psi}({\bf x}_{n,1},t) \hat{U}(t) 
 \cdots
\hat{U}^\dagger(t) \hat{\psi}({\bf x}_{n,n},t) \hat{U}(t)   |\Psi(t_0)\rangle .  
\end{equation}
Using $\hat{U}(t)\hat{U}^\dagger(t)=1$ and
\begin{equation}\label{nikolic:e3.52}
 \hat{U}(t)   |\Psi(t_0)\rangle = |\Psi(t)\rangle , \;\;\;\;
\hat{U}(t)   |0(t_0)\rangle = |0(t)\rangle ,
\end{equation}
the expression further simplifies 
\begin{equation}\label{nikolic:e3.53}
 \tilde{\psi}_n({\bf x}_{n,1},\ldots,{\bf x}_{n,n};t) = 
 d_n
\langle 0(t)|  \hat{\psi}({\bf x}_{n,1},t)  \cdots
\hat{\psi}({\bf x}_{n,n},t)    |\Psi(t)\rangle . 
\end{equation}
In practical applications of QFT in particle physics, one usually calculates the 
$S$-matrix, corresponding to the limit $t_0\rightarrow -\infty$, 
$t\rightarrow\infty$. For Hamiltonians that conserve energy (such as (\ref{nikolic:e3.48}))
this limit provides the stability of the vacuum, i.e., obeys
\begin{equation}\label{nikolic:e3.54}
\lim_{t_0\rightarrow -\infty, \; t\rightarrow\infty} \hat{U}(t)   |0(t_0)\rangle = 
e^{-i\varphi_0} |0(t_0)\rangle , 
\end{equation}
where $\varphi_0$ is some physically irrelevant phase \cite{bd2}. 
Essentially, this is because the integrals of the type 
$\int_{-\infty}^{\infty} dt' \cdots$ produce $\delta$-functions
that correspond to energy conservation, so the vacuum remains stable
because particle creation from the vacuum would violate energy conservation.
Thus we have
\begin{equation}\label{nikolic:e3.55}
 |0(\infty)\rangle =e^{-i\varphi_0}|0(-\infty)\rangle \equiv  e^{-i\varphi_0}|0\rangle .
\end{equation}
The state
\begin{equation}\label{nikolic:e3.55.1}
 |\Psi(\infty)\rangle=\hat{U}(\infty)|\Psi(-\infty)\rangle 
\end{equation}
is not trivial, but whatever it is, it has some expansion of the form
\begin{equation}\label{nikolic:e3.56}
 |\Psi(\infty)\rangle =  \sum_{n=0}^{\infty} c_{n}(\infty)|\Psi_{n}\rangle , 
\end{equation}
where $c_{n}(\infty)$ are some coefficients. 
Plugging (\ref{nikolic:e3.55}) and (\ref{nikolic:e3.56})
into (\ref{nikolic:e3.53}) and recalling (\ref{nikolic:e3.10})-(\ref{nikolic:e3.12.1}),
we finally obtain
\begin{equation}\label{nikolic:e3.57}
 \tilde{\psi}_n({\bf x}_{n,1},\ldots,{\bf x}_{n,n};\infty)=
e^{i\varphi_0} c_n(\infty) \psi_n({\bf x}_{n,1},\ldots,{\bf x}_{n,n};\infty) .
\end{equation}
This demonstrates the consistency of (\ref{nikolic:s2'}), because (\ref{nikolic:e3.55.1}) 
should be recognized as the standard description of evolution from
$t_0\rightarrow -\infty$ to $t\rightarrow \infty$ (see, e.g., \cite{chengli,bd2}),
showing that the coefficients $c_n(\infty)$ are the same as those described 
by standard $S$-matrix theory in QFT.
In other words, (\ref{nikolic:s2'}) is a natural many-time generalization of the concept of 
single-time evolution in interacting QFT.

\subsection{Generalization to other types of particles}
\label{nikolic:SEC3.4}

In Secs.~\ref{nikolic:SEC3.2} and \ref{nikolic:SEC3.3} we have discussed in detail
scalar hermitian fields, corresponding to spinless uncharged particles.
In this subsection we briefly discuss how these results can be
generalized to any type of fields and the corresponding particles.

In general, fields $\phi$ carry some additional labels which we 
collectively denote by $l$, so we deal with fields $\phi_l$.
For example, spin 1 field carries a
polarization label (see Sec.~\ref{nikolic:SECspin1}), 
fermionic spin $\frac{1}{2}$ field carries a spinor index,
non-Abelian gauge fields carry internal indices of the gauge group, etc.
Thus Eq.~(\ref{nikolic:e3.10}) generalizes to
\begin{equation}\label{nikolic:e3.10gen}
  \hat{\psi}_{n,L_n}(x_{n,1}, \ldots , x_{n,n}) = 
 d_n 
S_{ \{x_{n,1}, \ldots , x_{n,n} \} } 
\hat{\psi}_{l_{n,1}}(x_{n,1}) \cdots \hat{\psi}_{l_{n,n}} (x_{n,n}) , 
\end{equation}
where $L_n$ is a collective label $L_n=(l_{n,1}, \ldots , l_{n,n} )$.
The symbol $S_{ \{x_{n,1}, \ldots , x_{n,n} \} }$ denotes symmetrization (antisymmetrization) over bosonic (fermionic) fields describing the same type of particles.
Hence, it is straightforward to make the appropriate generalizations of all results
of Secs.~\ref{nikolic:SEC3.2} and \ref{nikolic:SEC3.3}. For example, (\ref{nikolic:e3.27})
generalizes to 
\begin{equation}\label{nikolic:e3.27gen}
\Psi_{\vec{L}} (\vec{x}) = \sum_{n=0}^{\infty} \sum_{L_n} \tilde{\Psi}_{n,L_n}
(\vec{x}) ,
\end{equation}
with self-explaining notation.

To further simplify the notation, we introduce the column 
$\Psi\equiv \{\Psi_{\vec{L}} \}$ and the row
$\Psi^{\dagger}\equiv \{\Psi^*_{\vec{L}} \}$.
With this notation, the appropriate generalization of (\ref{nikolic:e3.26.1}) can be written as
\begin{equation}\label{nikolic:e3.26.1gen}
 \int {\cal D}\vec{x} \, \sum_{{\vec{L}}} \Psi^*_{\vec{L}}(\vec{x})
\Psi_{\vec{L}}(\vec{x}) \equiv \int {\cal D}\vec{x} \,
\Psi^{\dagger}(\vec{x}) \Psi(\vec{x})
=1 .
\end{equation}

\subsection{Probabilistic interpretation}

The quantity
\begin{equation}\label{nikolic:e4.1}
 {\cal D}P=\Psi^{\dagger}(\vec{x}) \Psi(\vec{x}) \, {\cal D}\vec{x}
\end{equation}
is naturally interpreted as the probability of finding the system in the 
(infinitesimal) configuration-space volume ${\cal D}\vec{x}$ around a 
point $\vec{x}$ in the configuration space. Indeed, such an interpretation
is consistent with our normalization conditions such as 
(\ref{nikolic:e3.26.1}) and (\ref{nikolic:e3.26.1gen}). In more physical terms, 
(\ref{nikolic:e4.1}) gives the 
joint probability that the particle $1$ is found at the spacetime position 
$x_1$,  particle $2$ at the spacetime position $x_2$, etc.

As a special case, consider an $n$-particle state
$\Psi(\vec{x})=\Psi_n(\vec{x})$. It really depends only on
$n$ spacetime positions $x_{n,1},\ldots x_{n,n}$. With respect to
all other positions $x_B$, $\Psi$ is a constant. Thus, the probability
of various positions  $x_B$ does not depend on $x_B$; such a particle can be found
anywhere and anytime with equal probabilities. There is an infinite number of such
particles. Nevertheless, the Fourier transform of such a wave function reveals
that the 4-momentum $k_B$ of these particles is necessarily zero; they have neither
3-momentum nor energy. For that reason, such particles can be thought of as ``vacuum'' particles. 
In this picture, an $n$-particle state $\Psi_n$ is thought of as a state describing
$n$ ``real'' particles and an infinite number of ``vacuum'' particles.

To avoid a possible confusion with the usual notions of vacuum  
and real particles in QFT, in the rest of the paper
we refer to ``vacuum'' particles as {\it dead} particles
and ``real'' particles as {\it live} particles. Or let us be 
slightly more precise: We say that the 
particle $A$ is dead if the wave function in the momentum space
$\Psi(\vec{k})$ vanishes for all values of $k_A$ except $k_A=0$.
Similarly, we say that the particle $A$ is live if it is not dead.

The properties of live particles associated with the state $\Psi_n(\vec{x})$ can also be 
represented by the wave function $\psi_n(x_{n,1},\ldots,x_{n,n} )$. By averaging over
physically uninteresting dead particles, (\ref{nikolic:e4.1}) reduces to
 \begin{equation}\label{nikolic:e4.2}
 dP  =  \psi_n^{\dagger}(x_{n,1},\ldots,x_{n,n}) \psi_n(x_{n,1},\ldots,x_{n,n})
 \, d^4x_{n,1}\cdots d^4x_{n,n},
\end{equation}
which involves only live particles. In this way, the probabilistic 
interpretation is reduced to the probabilistic interpretation of relativistic
QM with a fixed number of particles, which is studied in Sec.~\ref{nikolic:RQ1}.

Now let us see how the wave functions representing the states in interacting QFT are interpreted
probabilistically. Consider the wave function $\tilde{\psi}_n(x_{n,1},\ldots,x_{n,n})$
given by (\ref{nikolic:s2'}). For example, it may vanish for small values of 
$x^0_{n,1}, \dots, x^0_{n,n}$, but it may not vanish for their large values. Physically, it means
that these particles cannot be detected in the far past (the probability is zero), but that
they can be detected in the far future. This is nothing but a probabilistic description of
the creation of $n$ particles that have not existed in the far past. Indeed, 
the results obtained in Sec.~\ref{nikolic:SEC3.3} (see, in particular, (\ref{nikolic:e3.57}))
show that such probabilities are consistent with the probabilities of particle creation obtained
by the standard $S$-matrix methods in QFT.

Having developed the probabilistic interpretation, we can also calculate the average values
of various quantities. In particular, the average value of the 4-momentum
$P^{\mu}_A$ is
\begin{equation}\label{nikolic:e4.6}
 \langle P^{\mu}_A \rangle= \int {\cal D}\vec{x} \, \Psi^{\dagger}(\vec{x}) \hat{P}^{\mu}_A
\Psi(\vec{x}) ,
\end{equation}
where $\hat{P}^{\mu}_A=i\partial^{\mu}_A$ is the 4-momentum operator.
Eq. (\ref{nikolic:e4.6})  can also be written as
\begin{equation}\label{nikolic:e4.9}
 \langle P^{\mu}_A \rangle= \int {\cal D}\vec{x}\, \rho(\vec{x}) U^{\mu}_A(\vec{x}) ,
\end{equation}
where
\begin{equation}\label{nikolic:e4.10}
\rho(\vec{x})=\Psi^{\dagger}(\vec{x})\Psi(\vec{x})
\end{equation}
is the probability density and
\begin{equation}\label{nikolic:e4.11}
 U^{\mu}_A(\vec{x})=\frac{J^{\mu}_A(\vec{x})}{\Psi^{\dagger}(\vec{x})\Psi(\vec{x})} .
\end{equation}
Here $J^{\mu}_A$ is given by an obvious generalization of (\ref{nikolic:e3.32}) 
\begin{equation}\label{nikolic:e4.12}
 J^{\mu}_A(\vec{x})=
\frac{i}{2} \Psi^{\dagger}(\vec{x})\!\stackrel{\leftrightarrow\;}{\partial^{\mu}_A}\! 
\Psi (\vec{x}) .
\end{equation}
The expression (\ref{nikolic:e4.9}) will play an important role in the next subsection.

\subsection{Bohmian interpretation}

In the Bohmian interpretation, each particle has some trajectory
$X^{\mu}_A(s)$.
Such trajectories must be consistent with the probabilistic interpretation (\ref{nikolic:e4.1}).
Thus, we need a velocity function $V^{\mu}_A(\vec{x})$, so that the trajectories
satisfy
\begin{equation}\label{nikolic:e4.13}
 \frac{dX^{\mu}_A(s)}{ds}=V^{\mu}_A(\vec{X}(s)) ,
\end{equation}
where the velocity function must be such that the following conservation equation is obeyed
\begin{equation}\label{nikolic:e4.14}
 \frac{\partial \rho(\vec{x})}{\partial s} + 
\sum_{A=1}^{\infty}\partial_{A\mu}[\rho(\vec{x}) V^{\mu}_A(\vec{x}) ] =0.
\end{equation}
Namely, if a statistical ensemble of particle positions in spacetime has the distribution 
(\ref{nikolic:e4.10}) for some initial $s$, then (\ref{nikolic:e4.13}) and (\ref{nikolic:e4.14}) will 
provide that this statistical ensemble will also have the distribution 
(\ref{nikolic:e4.10}) for {\em any} $s$, making the trajectories consistent with (\ref{nikolic:e4.1}).
The first term in (\ref{nikolic:e4.14}) trivially vanishes: $ \partial \rho(\vec{x})/\partial s =0$.
Thus, the condition (\ref{nikolic:e4.14}) reduces to the requirement
\begin{equation}\label{nikolic:e4.15} 
\sum_{A=1}^{\infty}\partial_{A\mu}[\rho(\vec{x}) V^{\mu}_A(\vec{x}) ] =0.
\end{equation}
In addition, we require that the average velocity should be proportional to the average momentum
(\ref{nikolic:e4.9}), i.e., 
\begin{equation}\label{nikolic:e4.16}
\int {\cal D}\vec{x}\, \rho(\vec{x}) V^{\mu}_A(\vec{x}) 
 = {\rm const} \times \int {\cal D}\vec{x}\, \rho(\vec{x}) U^{\mu}_A(\vec{x}) .
\end{equation} 
In fact, the constant in (\ref{nikolic:e4.16}) is physically irrelevant, because it can always be
absorbed into a rescaling of the parameter $s$ in (\ref{nikolic:e4.13}). Thus
we fix ${\rm const}=1$.

As a first guess, Eq.~(\ref{nikolic:e4.16}) with  ${\rm const}=1$ suggests that one could take
$V^{\mu}_A=U^{\mu}_A$. However, it does not work in general. Namely, 
from (\ref{nikolic:e4.10}) and (\ref{nikolic:e4.11}) we see that 
$\rho U^{\mu}_A =J^{\mu}_A$, and we have seen in (\ref{nikolic:e3.33})
that $ J^{\mu}_A$ does not need to be conserved. Instead, we have
\begin{equation}\label{nikolic:e4.18}
 \sum_{A=1}^{\infty}\partial_{A\mu}[\rho(\vec{x}) U^{\mu}_A(\vec{x})] = J(\vec{x}) ,
\end{equation}
where $J(\vec{x})$ is some function that can be calculated explicitly whenever 
$\Psi(\vec{x})$ is known. Therefore, instead of $V^{\mu}_A=U^{\mu}_A$
we must take
\begin{equation}\label{nikolic:e4.19}
 V^{\mu}_A(\vec{x})=U^{\mu}_A(\vec{x}) +
\rho^{-1}(\vec{x}) [e^{\mu}_A + E^{\mu}_A(\vec{x})] ,
\end{equation}
where
\begin{equation}\label{nikolic:e4.20}
 e^{\mu}_A=-{\cal V}^{-1} \int {\cal D}\vec{x}\, E^{\mu}_A(\vec{x}) ,
\end{equation}
\begin{equation}\label{nikolic:e4.21}
 E^{\mu}_A(\vec{x})=\partial^{\mu}_A 
\int {\cal D}\vec{x}'\, G(\vec{x},\vec{x}') J(\vec{x}') ,
\end{equation}
\begin{equation}\label{nikolic:e4.22}
 G(\vec{x},\vec{x}') = \int \frac{{\cal D}\vec{k}}{(2\pi)^{4\aleph_0}} 
\frac{e^{i\vec{k}(\vec{x}-\vec{x}')}}{\vec{k}^2} ,
\end{equation}
and $\aleph_0=\infty$ is the cardinal number of the set of natural numbers.
It is straightforward to show that
Eqs.~(\ref{nikolic:e4.21})-(\ref{nikolic:e4.22}) provide that (\ref{nikolic:e4.19}) obeys (\ref{nikolic:e4.15}), 
while (\ref{nikolic:e4.20}) provides that (\ref{nikolic:e4.19}) obeys (\ref{nikolic:e4.16}) with ${\rm const}=1$.

We note two important properties of (\ref{nikolic:e4.19}). First, if $J=0$ in (\ref{nikolic:e4.18}),
then $V^{\mu}_A=U^{\mu}_A$. In particular, since $J=0$ for free fields in states
with a definite number of particles (it can be derived for any type of particles 
analogously to the derivation of (\ref{nikolic:e3.34}) for spinless uncharged particles), 
it follows that $V^{\mu}_A=U^{\mu}_A$ for such states.
Second, if $\Psi(\vec{x})$ does not depend
on some coordinate $x^{\mu}_B$, then both $U^{\mu}_B=0$ and
$V^{\mu}_B=0$. [To show that $V^{\mu}_B=0$, note first that $J(\vec{x})$ defined
by (\ref{nikolic:e4.18}) does not depend on $x^{\mu}_B$ when $\Psi(\vec{x})$ does not depend
on $x^{\mu}_B$. Then the integration over $dx'^{\mu}_B$ in (\ref{nikolic:e4.21}) produces
$\delta(k^{\mu}_B)$, which kills the dependence on $x^{\mu}_B$ carried by 
(\ref{nikolic:e4.22})].
This implies that dead particles have zero 4-velocity.

Having established the general theory of particle trajectories by the results above, 
now we can discuss particular consequences. 

The trajectories are determined uniquely if the initial spacetime positions
$X^{\mu}_A(0)$ in (\ref{nikolic:e4.13}), for all $\mu=0,1,2,3$, $A=1,\ldots, \infty$,
are specified. In particular,
since dead particles have zero 4-velocity, such particles do not really have trajectories
in spacetime. Instead, they are represented by dots in spacetime.
The spacetime positions of these dots are 
specified by their initial spacetime positions.

Since $\rho(\vec{x})$ describes probabilities for particle creation and destruction, 
and since (\ref{nikolic:e4.14}) provides that particle trajectories are such that 
spacetime positions of particles
are distributed according to $\rho(\vec{x})$, it implies that particle trajectories are
also consistent with particle creation and destruction. In particular, the trajectories
in spacetime may have beginning and ending points, which correspond to points
at which their 4-velocities vanish. For example,
the 4-velocity of the particle A vanishes if the conditional wave function
$\Psi(x_A,\vec{X}')$ does not depend on $x_A$ (where $\vec{X}'$ denotes
the actual spacetime positions of all particles except the particle $A$).

One very efficient mechanism of destroying particles is through the interaction
with the environment, such that the total quantum state takes the form 
(\ref{nikolic:e3.4}). The environment wave functions
$(\vec{x}|E_0\rangle$, $(\vec{x}|E_{n;b_1,\ldots,b_n} \rangle$ do not overlap,
so the particles describing the environment can be in the support of only one of these
environment wave functions. Consequently, the conditional wave function
is described by only one of the terms in the sum (\ref{nikolic:e3.4}), which effectively
collapses the wave function to only one of the terms in (\ref{nikolic:e3.3}). For example,
if the latter wave function is $(\vec{x}|b_1,\ldots,b_n \rangle$, then it depends on only
$n$ coordinates among all $x_A$. All other live particles from sectors with
$n'\neq n$ become dead, i.e., their 4-velocities become zero which appears as
their destruction in spacetime.
More generally, if the overlap between the environment wave functions is negligible 
but not exactly zero, then particles from sectors with
$n'\neq n$ will not become dead, but their influence on the environment will still be
negligible, which still provides an effective collapse to $(\vec{x}|b_1,\ldots,b_n \rangle$.

Another physically interesting situation is when the entanglement with the environment
takes the form (\ref{nikolic:e3.6}), where $|\xi\rangle$ are coherent states. 
In this case, the behavior of the environment can very well be described in terms 
of an environment that responds to a presence of classical fields.
This explains how classical fields may appear at the
macroscopic level, even though the microscopic ontology is described
in terms of particles. 
Since $|\xi\rangle$ is a superposition of states with all possible numbers of particles,
trajectories of particles from sectors with different numbers of particles coexist;
there is an infinite number of live particle trajectories in that case. 

\section{Conclusion}

The usual formulation of Bohmian mechanics is not relativistic covariant because 
it is based on standard QM which is also not relativistic covariant.
Thus, to make  Bohmian mechanics covariant, one needs
first to reformulate the standard QM in a covariant way, such that
time is treated on an equal footing with space. More specifically, it means the following. First, 
the space probability density should be generalized to the spacetime probability density.
Second, the single-time wave function should be generalized to the many-time wave function.
When standard QM is generalized in that way, then the construction
of a relativistic-covariant version of Bohmian mechanics is straightforward.

To make the Bohmian mechanics of particles compatible with QFT and particle creation and destruction,
one needs to do the following. First, QFT states should be represented by wave functions
that depend on an infinite number of coordinates. Second, one needs to
use the quantum theory of measurements, which then leads to
an effective collapse into states of definite number of particles.

\section*{Acknowledgements}
\addtocontents{toc}{\protect\vspace*{\protect\baselineskip}}
\addcontentsline{toc}{section}{\slshape Acknowledgements}

This work was supported by the Ministry of Science of the
Republic of Croatia under Contract No.~098-0982930-2864.


\end{document}